\documentclass[ijoc,nonblindrev]{mystyle}
\OneAndAHalfSpacedXII % current default line spacing
\usepackage[font=scriptsize]{caption}
\usepackage{anyfontsize}
\usepackage{subcaption}
\usepackage{graphicx}
\usepackage{times,scalefnt}
\usepackage{amsmath,amssymb,mathtools}
\usepackage{tikz}
\usetikzlibrary{matrix,decorations.pathreplacing, calc, positioning, shapes.callouts, shapes.geometric}
\usepackage[boxruled,vlined,linesnumbered,commentsnumbered]{algorithm2e}

\newcommand{\p}{{p}}
\newcommand{\y}{{y}}

\newcommand{\model}{{\eta}}

\newcommand{\thetav}{{\pmb{\theta}}}
\newcommand{\Sigmav}{{\sigma^2}}
\newcommand{\wb}{{\pmb{\eta}}}

\newcommand{\Kv}{{\mathbf{K}}}
\newcommand{\kv}{{\mathbf{k}}}

\newcommand{\rhov}{{\pmb{\zeta}}}
\newcommand{\kb}{{\mathbf{k}}} 
\newcommand{\Kb}{{\mathbf{K}}} 

\def\pdfN#1#2#3{f_{\mathcal{N}}\left(#1; #2, #3 \right)}

\usepackage{natbib}
 \bibpunct[, ]{(}{)}{,}{a}{}{,}%
 \def\bibfont{\small}%
\TheoremsNumberedThrough     % Preferred (Theorem 1, Lemma 1, Theorem 2)
\EquationsNumberedThrough    % Default: (1), (2), ...
\begin{document}
%%%%%%%%%%%%%%%%

% Outcomment only when entries are known. Otherwise leave as is and 
%   default values will be used.
%\setcounter{page}{1}
%\VOLUME{00}%
%\NO{0}%
%\MONTH{Xxxxx}% (month or a similar seasonal id)
%\YEAR{0000}% e.g., 2005
%\FIRSTPAGE{000}%
%\LASTPAGE{000}%
%\SHORTYEAR{00}% shortened year (two-digit)
%\ISSUE{0000} %
%\LONGFIRSTPAGE{0001} %
%\DOI{10.1287/xxxx.0000.0000}%

% Author's names for the running heads
% Sample depending on the number of authors;
% \RUNAUTHOR{Jones}
% \RUNAUTHOR{Jones and Wilson}
% \RUNAUTHOR{Jones, Miller, and Wilson}
% \RUNAUTHOR{Jones et al.} % for four or more authors
% Enter authors following the given pattern:
\RUNAUTHOR{S\"{u}rer and Wild}

% Title or shortened title suitable for running heads. Sample:
% \RUNTITLE{Bundling Information Goods of Decreasing Value}
% Enter the (shortened) title:
\RUNTITLE{Performance Analysis for Parallel Calibration}

% Full title. Sample:
% \TITLE{Bundling Information Goods of Decreasing Value}
% Enter the full title:
\TITLE{Performance Analysis of Sequential Experimental Design for Calibration in Parallel Computing Environments}

% Block of authors and their affiliations starts here:
% NOTE: Authors with same affiliation, if the order of authors allows, 
%   should be entered in ONE field, separated by a comma. 
%   \EMAIL field can be repeated if more than one author
%SW: edit to put at top.
\ARTICLEAUTHORS{%
\AUTHOR{\"{O}zge S\"{u}rer}
\AFF{Department of Information Systems \& Analytics, Miami University, Oxford, Ohio, 45056, USA, \EMAIL{surero@miamioh.edu}, \URL{}}
\AUTHOR{Stefan M.\ Wild}
% SMW ORCID: 0000-0002-6099-2772
\AFF{Applied Math \& Computational Research, 
Lawrence Berkeley National Laboratory, Berkeley, California 94720, USA, \EMAIL{wild@lbl.gov};\\
Department of Industrial Engineering \& Management Sciences, Northwestern University, Evanston, Illinois 60208, USA \URL{}}
% Enter all authors
} % end of the block

\ABSTRACT{%
The unknown parameters of simulation models often need to be calibrated using observed data. When simulation models are expensive, calibration is usually carried out with an emulator. The effectiveness of the calibration process can be significantly improved by using a sequential selection of parameters to build an emulator. The expansion of parallel computing environments--from multicore personal computers to many-node servers to large-scale cloud computing environments--can lead to further calibration efficiency gains by allowing for the evaluation of the simulation model at a batch of parameters in parallel in a sequential design. 
However, understanding the performance implications of different sequential approaches in parallel computing environments introduces new complexities since the rate of the speed-up is affected by many factors, such as the run time of a simulation model and the variability in the run time. 
This work proposes a new performance model to understand and benchmark the performance of different sequential procedures for the calibration of simulation models in parallel environments. We provide metrics and a suite of techniques for visualizing the numerical experiment results and demonstrate these with a novel sequential procedure. The proposed performance model, as well as the new sequential procedure and other state-of-art techniques, are implemented in the open-source Python software package Parallel Uncertainty Quantification (PUQ), which allows users to run a simulation model in parallel. 
%PUQ is an open-source software package at \url{https://github.com/parallelUQ/PUQ}.
%, released under MIT License. 
}%

% Sample 
%\KEYWORDS{deterministic inventory theory; infinite linear programming duality; 
%  existence of optimal policies; semi-Markov decision process; cyclic schedule}

% Fill in data. If unknown, outcomment the field
\KEYWORDS{statistical model calibration; computational benchmarking; algorithm comparison; active learning; software}

\maketitle
%%%%%%%%%%%%%%%%%%%%%%%%%%%%%%%%%%%%%%%%%%%%%%%%%%%%%%%%%%%%%%%%%%%%%%

% Samples of sectioning (and labeling) in IJOC
% NOTE: (1) \section and \subsection do NOT end with a period
%       (2) \subsubsection and lower need end punctuation
%       (3) capitalization is as shown (title style).
%
%\section{Introduction.}\label{intro} %%1.
%\subsection{Duality and the Classical EOQ Problem.}\label{class-EOQ} %% 1.1.
%\subsection{Outline.}\label{outline1} %% 1.2.
%\subsubsection{Cyclic Schedules for the General Deterministic SMDP.}
%  \label{cyclic-schedules} %% 1.2.1
%\section{Problem Description.}\label{problemdescription} %% 2.

% Text of your paper here
\section{Introduction}
\label{sec:intro}

% Start with information about calibration
Simulation models have been used for the study and analysis of complex systems' properties and behaviors in myriad engineering and science disciplines. Examples include the simulation of freeway traffic \citep{Chen2019}, the COVID-19 epidemic \citep{Yang2020}, storm surge \citep{Plumlee2021}, and nuclear physics \citep{Surer2022}.
In these and many other examples, the output of the simulation model depends on user-specified, but unknown parameters that affect the mechanics of the simulation.
Calibration is a way to infer the unknown parameters using data observed from the system of interest. This work focuses on Bayesian calibration, a specific approach to calibration that provides a framework for quantifying uncertainty in both model parameters and predictions of key quantities \citep{Sung2024}.
When the simulation model is computationally expensive to evaluate, calibration becomes a more challenging problem.
In this paper, we consider the situation where the simulation model is evaluated a limited number of times due to resource constraints (e.g., limited concurrent resources, a time or energy budget) and the expense of simulation run times.

In such a calibration setting, one typically fits a computationally cheaper emulator such as a Gaussian process (GP) \citep{Rasmussen2005, gramacy2020surrogates} in place of the simulation model, and then the calibration is performed with the emulator \citep{Higdon2004}. Consequently, calibration accuracy highly depends on the emulator's accuracy. In the basic case, the emulator is built with simulation outputs obtained by evaluating the simulation model at a fixed set of design points consisting of a set of parameters. Space-filling designs, such as Latin Hypercube Sampling (LHS) or Sobol sequences, are widely used to generate designs for general-purpose emulator construction; see \cite{santner2018design} for a detailed survey. The downside of space-filling designs for calibration is that when the parameter region of interest is small relative to the support of the parameter space, the simulation outputs are generated with design points far from the (unknown) region of interest. As a result, the emulator may not accurately predict the simulation output near the parameter values of interest, leading to inaccurate calibration results. Sequential design, also called active learning, can be used to provide more precise inferences through the use of information learned in the course of experimental design to select more relevant parameters for evaluation. In sequential design, a criterion to assess the value of evaluating the simulation model in a new set of parameters is derived from previous simulation evaluations. A criterion, also known as an acquisition function, is then optimized to select a new set of parameters. An acquisition function generally makes a tradeoff between exploitation and exploration. While exploitation aims to focus on identified regions of interest, exploration aims to discover the promising regions that have not yet been detected.

Sequential design has been studied to build emulators with high global prediction accuracy across the parameter space; see \cite{Sacks1989, Seo200, Wang2018, Binois2019, Sauer2022} for examples and \cite[Chapter~6]{gramacy2020surrogates} for an extensive review. However, these approaches often overlook the observed data and acquire parameters outside the region of interest, adding little value for parameter inference. Bayesian optimization (BO) is another commonly used sequential approach for solving optimization problems that involve expensive objective functions \citep{Frazier2018, Wang2020}. In BO, the probabilistic statistical model (typically a GP) is constructed to infer the predictive distribution of the function being optimized (i.e., the goodness-of-fit measure), rather than the predictive distribution of the simulation model output. Other approaches, such as those in \cite{Joseph2015, Kandasamy2015, Joseph2019, Jarvenpa2019, Fernandez2020}, focus on creating designs based on the emulation of a goodness-of-fit measure. However, these methods do not consider the valuable information embedded in the simulation model. Recently, there has been growing interest in sequential design procedures for calibration, aiming to better explore the region of interest by leveraging emulators of simulation models \citep{Damblin2018, Koermer2023, Surer2023, Surer2024}. Building on these recent studies, our work focuses on sequential procedures for calibration in parallel processing environments, utilizing emulators of simulation models.

As the number of alternative acquisition functions increases, an objective method for evaluating their performance becomes essential to select the most suitable function for a specific simulation model and computing paradigm. Users should first consider what aspect of the acquisition function is most important and decide on the calibration goal. Calibration goals can vary, such as finding parameters that minimize the distance between observed data and simulation output \citep{Tuo2015}, accurately predicting unseen data \citep{Koermer2023}, or better estimating the posterior density of parameters \citep{Surer2023}. Additionally, one might aim to reduce bias in the simulation output through calibration \citep{Morgan2022}. Identifying the right goal is crucial for determining the performance metrics that will be used to evaluate different experimental configurations and workflows.

Performance metrics are typically summarized across iterations in a fully sequential setting, where parameters are acquired one at a time. This implicitly measures the convergence of the acquisition function in terms of iteration count, assuming that the run time of each simulation evaluation is identical. However, the run time of the simulation model may vary across different subspaces of the parameter space, and the convergence rates and time complexity of acquisition functions can differ. In such cases, alternative performance metrics, such as wall-clock time or speed-up to achieve a certain level of calibration goal, can be considered for comparing acquisition functions.

\begin{figure}[ht]
\centering
    \begin{subfigure}{1\textwidth}
        \includegraphics[width=1\textwidth]{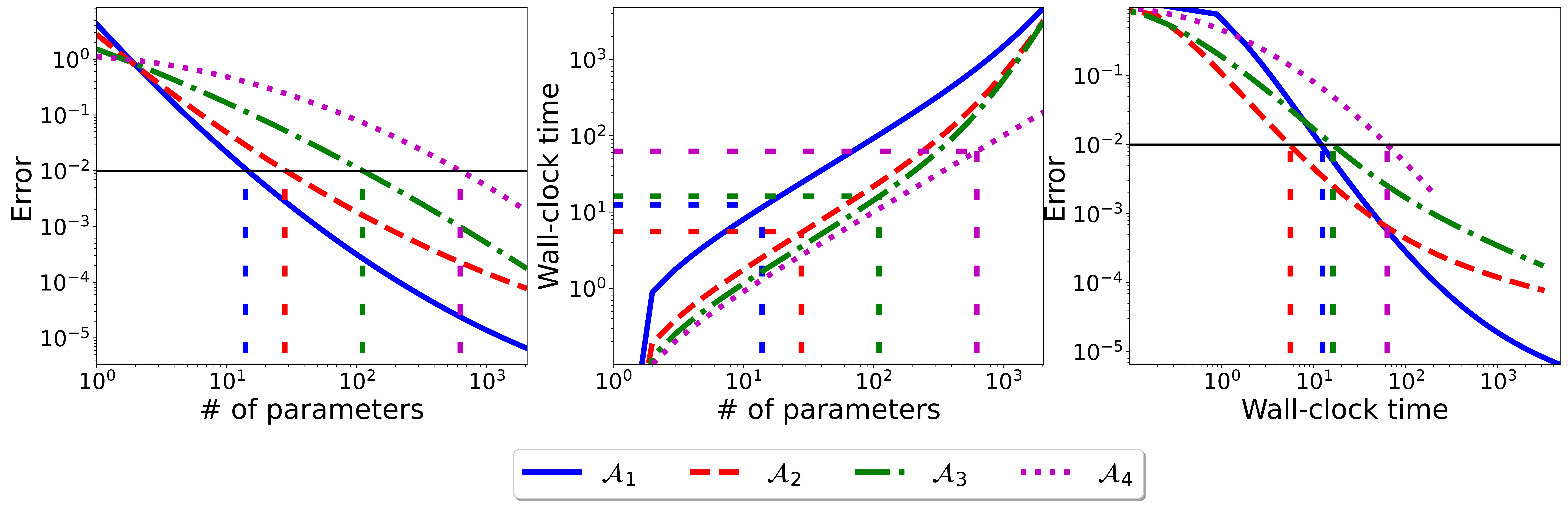}
    \end{subfigure}
    \caption{Illustration of the performance of four different acquisition functions $\mathcal{A}_1$, $\mathcal{A}_2$, $\mathcal{A}_3$, and $\mathcal{A}_4$. The left panel shows the calibration error of each occurrence when a simulation output is received from a set of acquired parameters. The middle panel shows the 
    total elapsed wall-clock time accounting for acquisition and simulation evaluation as the number of evaluated parameters grows.
 The right panel demonstrates the relationship between error and wall-clock time. The horizontal black line shows the desired error level of $10^{-2}$. }
    \label{fig:Figure1}
\end{figure}    
%wall-clock time to complete each simulation evaluation associated with each set of acquired parameters.
Figure~\ref{fig:Figure1} illustrates the performance of four acquisition functions denoted by $\mathcal{A}_1$, $\mathcal{A}_2$, $\mathcal{A}_3$, and $\mathcal{A}_4$ using a fully sequential procedure implemented on a test problem presented later in the paper. The left panel shows the calibration error for each instance where a simulation output is received from the acquired set of parameters. In other words, it visualizes the progress of each acquisition function with respect to the number of simulation evaluations. In the left panel, the error decays at different rates with each acquisition function. To reach the error level of $10^{-2}$ (illustrated with the black horizontal line), the required number of parameters is the smallest with the acquisition function $\mathcal{A}_1$ and largest with $\mathcal{A}_4$. The middle and right panels demonstrate the absolute speed of each acquisition function with the real (``wall clock'') time. At the error level of $10^{-2}$, although the acquisition function $\mathcal{A}_1$ requires less number of parameters as compared to $\mathcal{A}_2$, $\mathcal{A}_2$ reaches the desired objective earlier as illustrated in the middle and right panels since the time to acquire a single set of parameters is smaller with $\mathcal{A}_2$. On the other hand, to reach the error level of $10^{-4}$, the total elapsed wall-clock time, including both acquisition and simulation evaluation, is shorter with $\mathcal{A}_1$ since the smaller acquisition time with $\mathcal{A}_2$ is compensated with the faster error decay rate with $\mathcal{A}_1$. When comparing multiple acquisition functions, the choice of the desired accuracy level can favor slow-and-steady acquisition functions (by increasing the desired accuracy level) or rapid acquisition functions (by decreasing the desired accuracy level) depending on the run time of a simulation model.

The computational burden is further mitigated by evaluating the simulation model with a batch of parameters in parallel. Recent advances in parallel processing technology have helped reduce the total elapsed wall-clock time required to achieve a specific calibration goal. However, understanding the performance of an acquisition function for a given simulation model under different parallel settings remains a particularly complex task. This complexity arises from various computational factors, such as the convergence rate (and its variation with batch size), the time complexity of the acquisition function with different batch and parallelism sizes, and the characteristics of the simulation model (e.g., the nature of its dependence on the parameters, run time variability).

The main contributions of this paper address these challenges and include the following.
\begin{itemize}
    \item We provide a novel performance model to evaluate numerous configurations of sequential experiments in parallel processing environments. Our model helps practitioners and researchers assess how various computational factors impact the performance of sequential approaches, enabling more informed decisions in both research and practical applications.
    \item We propose and analyze a novel acquisition function for calibration. We also develop metrics and visualizations to understand various calibration goals.
    \item We analyze the speed-up behavior and scalability of various acquisition functions across different parallel configurations. 
    \item We conduct a comprehensive sensitivity analysis using our performance model to explore performance tradeoffs across various computational factors. Our analysis is motivated by our numerical analysis using different acquisition functions on a range of synthetic simulation models.
    \item We provide a toolset for practitioners to allocate their computational budget most efficiently and to make design decisions based on quantifiable performance metrics. Our toolset can be used as a benchmarking tool to provide a basis and guideline for deciding an appropriate configuration for future experiment designs.
    \item The performance model and the proposed sequential framework with the capability of parallel simulation evaluations are provided under the software library Parallel Uncertainty Quantification (PUQ) \citep{PUQpackage}, and installation instructions, source code, and the user manual can be found at \url{https://puq.readthedocs.io/en/latest/}. Moreover, our modular code framework allows researchers to create and implement novel sequential procedures, acquisition functions, and emulators in a parallel environment and compare them with the proposed approaches for different simulation models. This interchangeable and flexible code framework allows researchers to make new observations on the performance of sequential procedures for a variety of configurations, which, in turn, contributes to the development of new scientific methodologies.
\end{itemize}

The remainder of the paper is organized as follows. 
Section~\ref{sec:review} overviews the main steps of the sequential design procedure.
Section~\ref{sec:acquisition} introduces the proposed acquisition strategy. 
We then provide the performance model in Section~\ref{sec:performancemodel}. 
Finally, Section~\ref{sec:results} illustrates the computational and predictive advantages of the proposed approach using results from several simulation experiments and provides insights into the sequential procedure via the performance model.

\section{Background}
\label{sec:review}
Section~\ref{sec:seqdes} provides an overview of the sequential design, followed by a review of Bayesian calibration in Section~\ref{sec:calibration}. The GP model is then discussed in Section~\ref{sec:gp}. 

\subsection{Sequential Experimental Design}
\label{sec:seqdes}

\begin{algorithm}[t]
   \caption{Sequential experimental design in parallel processing environments \label{alg:Alg1}}

%\spacingset{1} % for alg only
    \emph{Initialize} $t=0$; $\mathcal{D}_t = \{(\thetav_i, \eta(\thetav_i)): i = 1, \ldots, n_0\}$ 
    
    \emph{Acquire} $w$ parameter sets such that $\{\thetav_{i} | i = 1, \ldots, w\}$ and submit the corresponding jobs
    
    \While {$n_t < n$} {
    
        $t \gets t + 1$, $\mathcal{D}_t \gets \mathcal{D}_{t-1}$

        \emph{Update} $\mathcal{D}_t$ with $b$ sets of parameters and simulation outputs; $n_t \gets n_{t-1} + b$
        
        \emph{Fit} an emulator with $\mathcal{D}_t$, $\widehat{\mathcal{D}}_t \gets \mathcal{D}_t$
    
        \For {$i = 1, \ldots, b$} {
            \emph{Generate} candidate solutions $\mathcal{L}_t$
            
            \emph{Select} $\displaystyle \thetav^{\mathrm{new}} \in \argmin\limits_{\thetav^* \in \mathcal{L}_t} \mathcal{A}_k(\thetav^*)|\widehat{\mathcal{D}}_t $
            
            \emph{Update} $\widehat{\mathcal{D}}_t \gets \widehat{\mathcal{D}}_t \cup (\thetav^{\rm new}, l)$ with a constant $l$ and the emulator with $\widehat{\mathcal{D}}_t$

            }
            
        Submit the corresponding $b$ jobs
                
        }
\end{algorithm}

The simulation model, represented with $\model$, is a function that takes a set of parameters $\thetav = (\theta_1, \ldots, \theta_p)^\top$ in a space $\Theta \subset \mathbb{R}^p$ as input and returns output $\model(\thetav)$. To effectively calibrate this model, Algorithm~\ref{alg:Alg1} presents the experimental design procedure, capable of performing both synchronous and asynchronous updates at each stage indexed by $t$. The parallel strategy follows a manager-worker computing paradigm using libEnsemble \citep{Hudson2022}. In this setup, a fixed set of $w$ workers performs simulation evaluations in parallel, while an additional worker (the manager) selects a batch of $b$ parameter sets to be evaluated next and assigns these $b$ jobs to available workers. In the synchronous update, the system waits for all $w$ workers to complete their jobs before acquiring $b=w$ new parameter sets. Conversely, the asynchronous update acquires $b < w$ new parameter sets as soon as $b$ workers become available. While the synchronous update can effectively reduce the total elapsed wall-clock time when simulation run times are roughly equal, it becomes less efficient for varying run times due to the long waiting periods, making the asynchronous approach more advantageous. In this setup, we consider workers to represent the number of concurrent simulations that can be run. However, workers can consist of various computational units, such as threads, cores, and nodes. Rather than assuming a one-to-one mapping between computational units and workers, one can consider parallelism within a simulation evaluation by using multiple computational units per worker. In this case, the run time for a single simulation decreases proportionally to the number of computational units due to parallelism. However, the largest possible batch size $b$ also decreases proportionally to the number of computational units.

Algorithm~\ref{alg:Alg1} begins with an initial design of size $n_0$. At each stage, $b$ simulation outputs are collected from workers (lines~4--6), and $b$ new sets of parameters are selected using the acquisition function $\mathcal{A}_k(\cdot)$ (lines~7--10). These new parameter sets are then assigned to available workers for simulation model evaluation (line~11). Selecting the best $b$ parameter sets at each stage creates a $b \times p$-dimensional optimization problem, which is generally more complex than solving $b$ separate $p$-dimensional optimization problems. To avoid this computational challenge, Algorithm~\ref{alg:Alg1} minimizes the acquisition function $b$ times over a candidate sets of parameters $\mathcal{L}_t$ (line~9). At each stage, the simulation data set $\mathcal{D}_t = \{(\thetav_i, \eta(\thetav_i)): i = 1, \ldots, n_t\}$ is utilized to build the emulator (Section~\ref{sec:gp}), which in turn is used to construct the acquisition function (Section~\ref{sec:acquisition}). When constructing a batch, since simulation outputs for new sets of parameters are not yet available, the emulator is updated with a constant $l$ as the simulation output (line~10). This approach, known as the constant liar strategy, is introduced in \cite{Ginsbourger2010} for BO. The algorithm terminates once $n$ simulation outputs are collected.

\subsection{Overview of Bayesian Calibration}
\label{sec:calibration}

We consider the data generation mechanism
\begin{equation}
    \y = \model(\thetav) + \epsilon, \quad \epsilon \sim {\rm N}(0, \Sigmav), \label{eq:statmodel}
\end{equation}
where $\epsilon$ represents the residual error. For simplicity, we assume $\Sigmav$ is known; if unknown, a plug-in estimator can be used. While many simulation models produce high-dimensional outputs (see \cite{Bayarri2007, Higdon2008, Conti2010, Huang2020} for examples), this paper focuses on scalar outputs from simulations. The selection of an acquisition function is influenced by the complexity of the response surface; for example, emulating high-dimensional outputs generally requires more model evaluations compared to scalar outputs to achieve the same level of prediction accuracy. Extending our acquisition function to handle high-dimensional outputs is left for future work, as the main focus here is on the computational aspects of the sequential procedure, rather than emulator complexities. Although we demonstrate the use of a GP for scalar outputs, our modular software framework is designed to accommodate different types of emulators, allowing for both scalar and high-dimensional outputs.

In the Bayesian calibration framework, model parameters are treated as random variables. The posterior $\p(\thetav|\y)$ represents the probability density of the parameter set $\thetav$ given the observed data $\y$, while the prior probability density $p(\thetav)$ captures the initial knowledge about $\thetav$ before any data is observed. The likelihood $p(\y|\thetav)$ quantifies how well a model output with a given set of parameters explains the observed data $\y$. According to Bayes' rule and the statistical model in \eqref{eq:statmodel}, the posterior density is given by
\begin{align} \label{eq:posterior}
    \begin{split}
    p(\thetav|\y) = \frac{ p(\y|\thetav) p(\thetav)}{\int_{\Theta}  p(\y|\thetav') p(\thetav') {\rm d \thetav'}} & \propto \tilde{p}(\thetav|\y) =  p(\y|\thetav) p(\thetav) = \frac{1}{\sqrt{2 \pi \sigma^2}} \exp\left(-\frac{(y - \eta(\thetav))^2}{2\sigma^2}\right) p(\thetav),
    \end{split}
\end{align}
where $\tilde{p}(\thetav|\y)$ represents the unnormalized posterior, commonly used in Bayesian calibration with Markov chain Monte Carlo (MCMC) methods to compute the posterior density up to a normalization constant \citep{Gelman2004}. In \eqref{eq:posterior}, the simulation output $\eta(\thetav)$ is needed to evaluate the unnormalized posterior $\tilde{p}(\thetav|\y)$ at any $\thetav$. A GP emulator, as described in the following section, is commonly used as a cost-effective surrogate for the simulation output for computationally expensive simulations. This approach enables the approximation of the unnormalized posterior using the emulator.

\subsection{Gaussian Process Model}
\label{sec:gp}

At the end of stage $t$, the simulation data set $\mathcal{D}_t$ contains all the simulation data collected up to that stage. This dataset is then used to construct the GP emulator. Under a GP model, the simulation output at an unseen $\thetav$ is typically modeled as a zero-mean random process with a positive definite kernel function $k_t(\thetav, \thetav') = \tau_t^{2} c(\thetav, \thetav'; \rhov_t)$, where $\tau_t^{2}$ is a scaling parameter, $c(\thetav, \thetav'; \rhov_t)$ is the correlation function, and $\rhov_{t} = \left(\zeta_{t, 1}, \ldots, \zeta_{t, p}\right)^\top$ represents the lengthscale parameter. Given $\wb_t = \left(\eta(\thetav_1), \ldots, \eta(\thetav_{n_t})\right)^\top$, the GP model implies that $\eta(\thetav)|\wb_{t}$ follows a normal distribution with mean $m_{t}(\thetav)$ and variance $s^2_t(\thetav)$ such that
\begin{equation}
    \begin{aligned}
        & m_t(\thetav) = \kb_t^\top(\thetav) \Kb_t^{-1} \wb_t \quad \text{and} \quad s^2_t(\thetav) = k_t(\thetav, \thetav) - \kb^\top_t(\thetav) \Kb_t^{-1} \kb_t(\thetav). \label{eq:meanvar_latent}
    \end{aligned}
\end{equation}
Here, $\kv_t(\thetav) = \left(k_t\left(\thetav, \thetav_1\right), \ldots, k_t\left(\thetav, \thetav_{n_t}\right)\right)^\top$  consists of cross-kernel evaluations between $\thetav$ and $\thetav_i$, for $i=1, \ldots, n_t$. The matrix $\Kv_t$ is an $n_t \times n_t$ matrix where the $(i,j)$-th entry is $k_t(\thetav_i, \thetav_j) + \upsilon_t \delta_{i=j}$ for $1 \leq i, j \leq n_t$. In addition, $\upsilon_t > 0$ represents a nugget parameter, and $\delta_{i=j}$ is the Kronecker delta function, which equals 1 when $i=j$ and 0 otherwise. 
For the correlation function, we use the separable version of the Mat\'ern correlation function with smoothness parameter 1.5 \citep{Rasmussen2005} such that
\begin{equation}
    \begin{aligned}
        c(\thetav, \thetav'; \rhov_t) = & \bigg[\prod_{l=1}^p (1 + |(\theta_l - \theta_l')\exp(\zeta_{t,l})|)\bigg] \times  \exp\bigg(-\sum_{l=1}^p |(\theta_l - \theta_l')\exp(\zeta_{t,l})|\bigg).
    \end{aligned}
\end{equation}

\section{Acquisition Functions}
\label{sec:acquisition}

In Algorithm~\ref{alg:Alg1}, the acquisition function $\mathcal{A}_k(\cdot)$ guides the choice of the next evaluation point for the simulation model based on the current data $\mathcal{D}_t$. In this section, we introduce two novel acquisition functions for calibration—the probability of improvement (PI) and expected improvement (EI)—adapted from BO. EI/PI selects parameters with the highest expectation or probability of improving upon the best value achieved so far  \citep{Jones1998}. However, the main limitation of EI/PI  is its tendency to prioritize exploitation, which can lead to getting trapped in local optima, thereby limiting exploration \citep{Mohammadi2022}. To address this,  we employ a variance-based criterion, the expected integrated variance (EIVAR) \citep{Surer2023}, which facilitates the exploration of high posterior regions. To balance exploitation and exploration, we propose alternating between EI/PI and EIVAR criteria at each stage, a strategy we refer to as HYBRID in our experiments. The benchmarks in Section~\ref{sec:results} employ EI/PI, EIVAR, and HYBRID, with the resulting experimental data forming the basis for the sensitivity analysis using the performance model introduced in Section~\ref{sec:performancemodel}.

We derive the EI/PI criterion based on the absolute difference between the simulation output and the observed data, as parameter sets that minimize absolute loss typically reside in regions of high posterior density. The use of EI/PI with absolute loss has been previously applied in Bayesian history matching to exclude implausible regions \citep{Gardner2019}. In this work, EI/PI is adapted to focus specifically on high posterior regions. Let $\delta_t$ represent the minimum loss obtained up to stage $t$, defined as $\delta_t = \min\{\left|\y - \model(\thetav_j)\right|, j = 1, \ldots, n_t\}$. At each stage, PI identifies a new set of parameters for which the difference between the observed data and its simulation output is likely to fall within the current error bounds $\left[-\delta_t, \delta_t\right]$. The PI at any $\thetav^*$ is then calculated by $\mathbb{E}_{\model(\thetav^*)| \mathcal{D}_{t}} \left[\mathbb{P}\left(\left|\epsilon^*\right| \leq \delta_t\right)\right]$, where $\epsilon^*$ is the random difference between the (unknown) simulation output $\model(\thetav^*)$ and the observed data $y$. The PI criterion is derived in Appendix~\ref{app:proofs} and is given as follows.
\begin{proposition}\label{prop:expectedprobimprovement}
The expected probability of improvement (PI) for $\thetav^*$ is given by
    \begin{align} \label{eq:expectedprobimprovement}
        \begin{split}
            \Phi\left(\frac{\delta_t - \left(y - m_t(\thetav^*)\right)}{\sqrt{\sigma^2 + s_t^2\left(\thetav^*\right)}}\right) -
            \Phi\left(\frac{- \delta_t - \left(y - m_t(\thetav^*)\right)}{\sqrt{\sigma^2 + s_t^2\left(\thetav^*\right)}}\right),
        \end{split}
    \end{align}
where $\delta_t = \min\{\left|\y - \model(\thetav_j)\right|, j = 1, \ldots, n_t\}$ and $\Phi$ represents the cumulative distribution function (CDF) of a standard normal random variable.
\end{proposition}

To develop the EI criterion, we consider its complement, the expected unimprovement, defined as follows
\begin{align} \label{acquisitionfuncEI}
    \begin{split}
        & \mathbb{E}_{\model(\thetav^*)| \mathcal{D}_{t}}\left[\max\left(\epsilon^* - \delta_t, 0\right)\right] + \mathbb{E}_{\model(\thetav^*)|\mathcal{D}_{t}}\left[\max\left(-\epsilon^* - \delta_t, 0\right)\right].
    \end{split}
\end{align}
Unimprovement occurs if the difference between the observed data and the simulation output, denoted as $\epsilon^*$, exceeds the absolute error $\delta_t$ or falls below its negative, $-\delta_t$. Therefore, at each stage, the goal is to identify the set of parameters that minimizes \eqref{acquisitionfuncEI}, or equivalently, maximizes the negative of \eqref{acquisitionfuncEI}, which we refer to as EI throughout the paper. We derive the EI as follows, with a detailed proof provided in Appendix~\ref{app:proofs}.
\begin{proposition}\label{prop:expectedimprovement}
The expected unimprovement for $\thetav^*$ is given by 
    \begin{align} \label{eq:expectedimprovement}
        \begin{split}
            &\left(\y - m_t(\thetav^*) - \delta_t\right) \left(1 - \Phi\left(\frac{\delta_t - \left(\y - m_t(\thetav^*)\right)}{\sqrt{\sigma^2 + s_t^2\left(\thetav^*\right)}}\right)\right) + \sqrt{\sigma^2 + s_t^2\left(\thetav^*\right)} \phi\left({\frac{\delta_t -\left( \y - m_t(\thetav^*)\right)}{\sqrt{\sigma^2 + s_t^2\left(\thetav^*\right)}}}\right) \\
            & + \left( - \delta_t - \left(\y - m_t(\thetav^*)\right)\right) \Phi\left(\frac{-\delta_t - \left(\y - m_t(\thetav^*)\right)}{\sqrt{\sigma^2 + s_t^2\left(\thetav^*\right)}}\right) + \sqrt{\sigma^2 + s_t^2\left(\thetav^*\right)}\phi\left({\frac{-\delta_t - \left(\y - m_t(\thetav^*)\right)}{\sqrt{\sigma^2 + s_t^2\left(\thetav^*\right)}}}\right),
        \end{split}
    \end{align}
where $\delta_t = \min\{\left|\y - \model(\thetav_j)\right|, j = 1, \ldots, n_t\}$ and $\Phi$ and $\phi$ represent the CDF and the probability density function (PDF) of a standard normal random variable, respectively.
\end{proposition}

\begin{figure}[ht]
    \begin{subfigure}{0.45\textwidth}
        \includegraphics[width=1\textwidth]{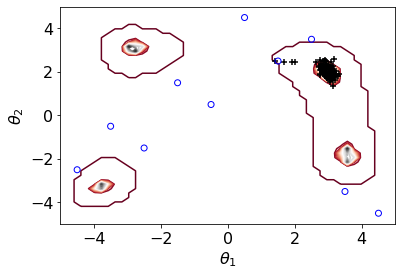}
    \end{subfigure}
    \begin{subfigure}{0.45\textwidth}
        \includegraphics[width=1\textwidth]{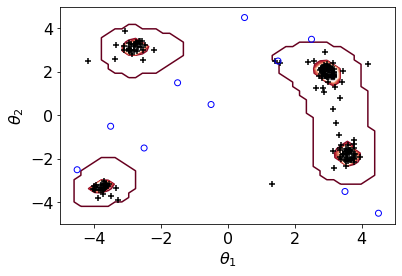}
    \end{subfigure}
    \caption{190 sets of acquired parameters (plus markers) using EI criterion (left) and the proposed HYBRID approach (right). Blue circles illustrate $n_0 = 10$ samples to initiate the procedure. The lines represent the level curves of the posterior.}
    \label{fig:Figure2}
\end{figure}
As noted, the main limitation of EI/PI is their tendency to become trapped in local optima due to their focus on exploitation, which can be particularly problematic when there are multiple high posterior regions. To overcome this limitation, we employ the variance-based criterion EIVAR \citep{Surer2023},  which accounts for the total uncertainty in the posterior density and thus improves the exploration of high posterior regions. We calculate the EIVAR at any $\thetav^*$ criterion as follows
\begin{align} \label{acquisitionfuncEIVAR}
    \begin{split}
        & \int_{\Theta} \mathbb{E}_{\model(\thetav^*) | \mathcal{D}_{t}} \left(\mathbb{V} \left[\tilde{p}(\thetav|\y)| \left(\thetav^*, \model(\thetav^*)\right) \cup \mathcal{D}_{t} \right] \right) d \thetav.
    \end{split}
\end{align}
EIVAR selects the parameter that minimizes the aggregated unnormalized posterior variance across the parameter space as if the parameter and its unknown output were included in the simulation data.  
The derivation of EIVAR follows from \cite{Surer2023} and it is approximated as
\begin{align} \label{approximateEIV}
    \begin{split}
         & \frac{1}{|\Theta_{\rm ref}|} \sum_{\thetav \in \Theta_{\rm ref}} p(\thetav)^2 \left(\frac{f_\mathcal{N}\left(\y; \, m_{t}(\thetav), \, \frac{1}{2}\Sigmav + s_t^2(\thetav)\right)}{2\pi^{1/2}\sigma} \right. \\
        & \hspace{1.5in} - \left. \frac{f_\mathcal{N}\left(\y; \, m_{t}(\thetav), \, \frac{1}{2}\left(\Sigmav + s_t^2(\thetav) + \tau^2_t(\thetav, \thetav^*)\right)\right)}{2\pi^{1/2}|\Sigmav + s_t^2(\thetav) - \tau^2_t(\thetav, \thetav^*)|^{1/2}}\right),
    \end{split}
\end{align} 
where $\tau^2_{t}(\thetav, \thetav^*) = \text{cov}_{t}(\thetav, \thetav^*)^2/\left(s^2_{t}(\thetav^*) + \upsilon_t\right)$, $\text{cov}_{t}(\thetav, \thetav^*) = k_t(\thetav, \thetav^*) - \kv_t(\thetav)^\top \Kv_t^{-1}\kv_t(\thetav^*)$, and $\Theta_{\rm ref}$ is set of uniformly distributed reference parameters. $f_\mathcal{N}(\mathbf{a}; \, \mathbf{b}, \, \mathbf{c})$ denotes the probability density function of the normal distribution with mean $\mathbf{b}$ and variance $\mathbf{c}$, evaluated at the value $\mathbf{a}$.  

We propose a strategy, called HYBRID, which alternates between the EI/PI and EIVAR criteria at each stage to achieve a balance between exploitation and exploration. Figure~\ref{fig:Figure2} illustrates this approach by comparing 190 sets of acquired parameters using the EI criterion (left) and the HYBRID approach (right) for an example with two parameters (details are provided in Section~\ref{sec:results}). The EI criterion tends to overexploit the region around $(3.8, 2)$, sampling numerous parameters in this area. In contrast, the HYBRID approach alternates between EI and EIVAR at each stage, allowing EIVAR to explore previously undetected high posterior regions while EI focuses on regions already identified as high-posterior. This balanced approach helps avoid local optima and improves the overall sampling strategy.

\section{Performance Model Using Sequential Design}
\label{sec:performancemodel}

\begin{figure}[ht]
    \includegraphics[width=1\textwidth]{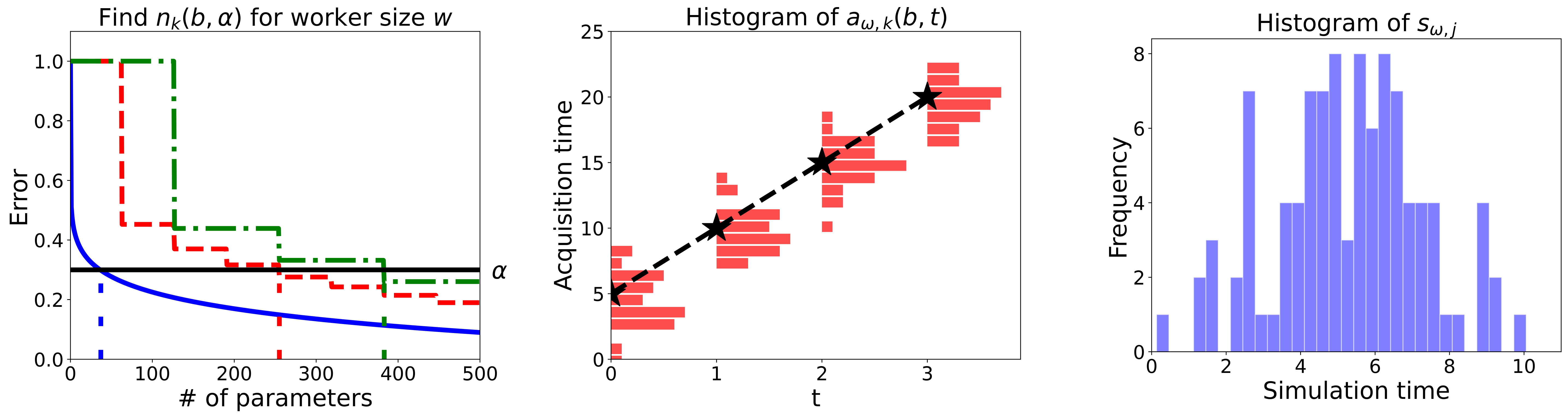}
    \caption{Illustration of the inputs utilized in the performance model in Algorithm~\ref{alg:Alg2}.}
    \label{fig:Figure3}
\end{figure}
Sequential procedures can offer more precise inferences compared to one-shot sampling methods while remaining computationally efficient. However, determining the best configuration and resource utilization is challenging because evaluating every possible configuration of the sequential procedure in Algorithm~\ref{alg:Alg1} for a computationally expensive simulation model is impractical due to time and computational limitations. To address this, we introduce a performance model in Algorithm~\ref{alg:Alg2} through Monte Carlo simulations. This model serves as a benchmarking tool, enabling sensitivity analysis of various computational factors and assisting users in fine-tuning aspects of experimental design, such as batch size $b$ and worker size $w$, to accommodate different acquisition functions with varying convergence rates and time complexities.

\begin{algorithm}[t]
   \caption{Performance model based on Monte Carlo simulation \label{alg:Alg2}}
    \textbf{Input}: batch size $b$, worker size $w$, acquisition function $\mathcal{A}_k$, $n_k(b, \alpha)$ \\ 
%\spacingset{1} % for alg only
\For {$ \omega \in \Omega$} {
    \emph{Initialize} $t=0$; $\mathcal{P}_{\omega,t}(w) = \{1, \ldots, w\}$; $n_t = w$
    
    $c_{\omega,j}^\mathcal{J}(b,w,k)  = s_{\omega,j} \quad \forall j=1, \ldots, w$; $c_{\omega,t}^\mathcal{S}(b,w,k) = 0$
    
    \While {$n_t < n_k(b, \alpha)$} {
    
        $t \gets t + 1$; $n_t \gets n_{t - 1}$; $\mathcal{P}_{\omega,t}(w) \gets \mathcal{P}_{\omega,t-1}(w)$
        
        $c_{\omega, t}^\mathcal{S}(b,w,k) = \max\left\{c_{\omega, t-1}^\mathcal{S}(b,w,k), {\left\{c_{\omega,j}^\mathcal{J}(b,w,k): j \in \mathcal{P}_{\omega,t}(w)\right\}}_{[b]}\right\} + a_{\omega,k}(b,t)$
           
        \emph{Remove} $b$ completed jobs from pending $\mathcal{P}_{\omega,t}(w)$ 
        
            \For {$i = 1, \ldots, b$} {
                
                \emph{Generate} job $n_t + i$
                
                $c^\mathcal{J}_{\omega, n_t + i}(b,w,k) = c_{\omega,t}^\mathcal{S}(b,w,k) + s_{\omega, n_t + i}$
                
                $\mathcal{P}_{\omega,t}(w) \gets \mathcal{P}_{\omega,t}(w) \cup \{n_t + i\}$
            }
        $n_t \gets n_t + b$
        }
    }
    \textbf{Output}: $\{c_{\omega, j}^{\mathcal{J}}(b,w,k): j = 1, \ldots, n_k(b, \alpha) \}$, $\{c_{\omega, t'}^{\mathcal{S}}(b,w,k): t' = 1, \ldots, t \}$
    
\end{algorithm}
Performance evaluation and benchmarking of algorithms have long been a focus in the optimization community \citep{Barr1993, Barr1992, McGeoch1996, Hall2007, Jorge2009}. Recently, \cite{Eckman2023} introduced diagnostic tools for assessing and comparing simulation-optimization solvers. While these algorithms are often used for parameter inference, our study shifts focus from solvers to comparing acquisition functions for designing simulation experiments for calibration. As highlighted by \cite{Kleijnen2005}, there is a need for more research and tools to identify experimental designs that best meet the needs and utilize them effectively. Driven by this need, our work emphasizes designing simulation experiments for calibration, considering the typical characteristics of simulation models and sequential data collection in a parallel environment where emulators act as proxies for simulation models. Building on the work of \cite{riche2012} on BO, we develop performance metrics and visualizations to analyze the behavior of algorithms under varying batch and worker sizes. We also provide a test bed of synthetic simulation models and acquisition functions to explore performance tradeoffs, considering both simulation evaluation and acquisition times.

Algorithm~\ref{alg:Alg2} presents pseudocode for a performance model with specified batch size $b$, worker size $w$, and acquisition function $\mathcal{A}_k$. Figure~\ref{fig:Figure3} provides a visual overview of the inputs required for this algorithm. We monitor progress over time and define $n_k(b, \alpha)$ as the total number of simulation outputs needed to achieve $\alpha$-level calibration accuracy with acquisition function $\mathcal{A}_k$ and batch size $b$. The $\alpha$-level corresponds to the highest accuracy, defined as the lowest calibration error throughout the paper. In deterministic optimization, progress curves show the evolution of solutions over time for a single solver on a given problem \citep{Eckman2023}. Similarly, we refer to plots illustrating the calibration error of an acquisition function after each simulation evaluation as progress curves. The metric, $n_k(b, \alpha)$, quantifies the simulation evaluations needed to solve a problem and is a key efficiency measure in optimization as well \citep{Beiranvand2017}. For instance, Figure~\ref{fig:Figure3} (left panel) displays three progress curves for different batch sizes, with the horizontal black line indicating the error level of $\alpha = 0.3$ and the vertical lines marking $n_k(b, \alpha)$ for three different batch sizes.  Algorithm~\ref{alg:Alg2} utilizes $n_k(b, \alpha)$ as an input to determine when to stop the Monte Carlo simulation. It is important to note that the characteristics of simulation models can greatly influence $n_k(b, \alpha)$, making it challenging to determine this value precisely without running the specific algorithm. Nevertheless, the examples in our PUQ software package can guide users in selecting an appropriate value for $n_k(b, \alpha)$.

In Algorithm~\ref{alg:Alg2}, due to the randomness in run times, the simulation results are replicated using different random values, with each replication indexed by $\omega \in \Omega$. For a specific replication $\omega$ (line~2), $b$ completed jobs are received (lines~7--8) and $b$ new parameters are acquired (lines~9--12) at each stage $t$. This process continues until $n_k(b, \alpha)$ parameters are acquired (line~5). The pending list $\mathcal{P}_{\omega,t}(w)$ keeps track of jobs that have been submitted to workers but not yet completed at stage $t$, and its size increases with worker size $w$. Each replication $\omega$ begins by submitting $w$ jobs to all available workers (line~3).

Besides $n_k(b, \alpha)$, the time required to acquire $b$ sets of parameters with acquisition function $\mathcal{A}_k$ at stage $t$, denoted by $a_{\omega,k}(b, t)$, is also an input for the performance model in Algorithm~\ref{alg:Alg2}, as illustrated in Figure~\ref{fig:Figure3} (middle panel). The acquisition time depends on both the batch size $b$ and the stage index $t$. For example, in Algorithm~\ref{alg:Alg1}, the time required to build an emulator increases with $t$ as the simulation data set grows. Similarly, larger batch sizes result in longer acquisition times due to the loop over $b$ in line~9. If available, varying acquisition times from numerical experiments should be used; otherwise, the mean acquisition time can be used as an input (dashed line in the middle panel). Furthermore, the time complexity of different acquisition functions varies. For instance, EI/PI requires less time than EIVAR, as EIVAR involves evaluating numerous probability density functions over a reference set of parameters to compute total uncertainty, making it more time-consuming. Moreover, their performance in parallel modes may differ based on batch size and how the sequential process is adapted for parallel execution. 
 
In addition to the acquisition times, the time required to complete a simulation evaluation for job $j$, denoted by $s_{\omega, j}$ (i.e., run time), is an input to the performance model (see the right panel in Figure~\ref{fig:Figure3}). Running a simulation model repeatedly with different parameter sets results in varying run times. In some cases, these differences are minimal, but in many applications, run times can vary significantly—either unusually long (possibly due to convergence issues) or exceptionally short (often due to failure) for various reasons. In parallel computing environments, accounting for both the evaluation times and their variability is crucial for fair performance analysis. In BO framework, \cite{Kandasamy18} model evaluation times using uniform, half-normal, and exponential distributions. However, in Algorithm~\ref{alg:Alg2}, since we do not assume any specific distribution, practitioners can use the run times of their model based on prior experiments and empirical distributions to better capture the full range of observed values in practice. 

We record the time $c_{\omega,j}^\mathcal{J}(b,w,k)$ at which each job $j$ ended and the end time $c_{\omega, t}^\mathcal{S}(b,w,k)$ for each stage $t$ within a given replication $\omega$. At each stage, the end times of jobs in the pending list $\mathcal{P}_{\omega,t}(w)$ are sorted in ascending order, and the $b$th job in this order is selected as the last job completed at the given stage (line~7). The acquisition process at stage $t$ begins on the manager worker once the previous stage concludes and the job in the $b$th order of the pending list $\mathcal{P}_{\omega,t}(w)$ is evaluated. Thus, the end time $c_{\omega, t}^\mathcal{S}(b,w,k)$ of stage $t$ is calculated as the sum of the acquisition start time at stage $t$ (i.e., $\max\left\{c_{\omega, t-1}^\mathcal{S}(b,w,k), {\left\{c_{\omega,j}^\mathcal{J}(b,w,k): j \in \mathcal{P}_{\omega,t}(w)\right\}}_{[b]}\right\}$) and the acquisition time $a_{\omega,k}(b,t)$ as in line~7. Once jobs are completed at the current stage, they are removed from the pending list (line~8). 

\section{Experiments}
\label{sec:results}

Section~\ref{sec:syntheticresults} considers the performance of the sequential procedure with acquisition functions described in Section~\ref{sec:acquisition} on various test functions. Motivated by the results in Section~\ref{sec:syntheticresults}, Section~\ref{sec:performanceexperiments} focuses on different computational factors of the sequential approach and examines their effect on the performance based on the model introduced in Section~\ref{sec:performancemodel}. We provide an open-source implementation of different acquisition functions and their parallel implementations in the \texttt{PUQ} Python package \citep{PUQpackage}. All the test functions are provided in the \texttt{examples} directory within the \texttt{PUQ} Python package.

\subsection{Simulations from Synthetic Probability Densities}
\label{sec:syntheticresults}

We evaluate the performance of the sequential experimental design using the proposed HYBRID approach, along with the EI and EIVAR acquisition functions. As a baseline, we also include a random sampling method from a uniform prior, referred to as RND. These different acquisition strategies are tested on six well-known synthetic test functions, each with distinct density profiles, as shown in Figure~\ref{synth_figs}. Detailed descriptions of each simulation model and the resulting likelihoods are provided in Appendix~\ref{app:setting}. In all cases, a uniform prior is employed, and an initial sample of $n_0 = 10$ points is drawn from this prior, with ranges specified in Appendix~\ref{app:setting}. At each stage, a candidate list $\mathcal{L}_t$ comprising 1,000 randomly sampled points from the uniform prior is generated for each synthetic function. Each experimental setting is replicated 30 times. 

\begin{figure}[ht]
        \includegraphics[width=1\textwidth]{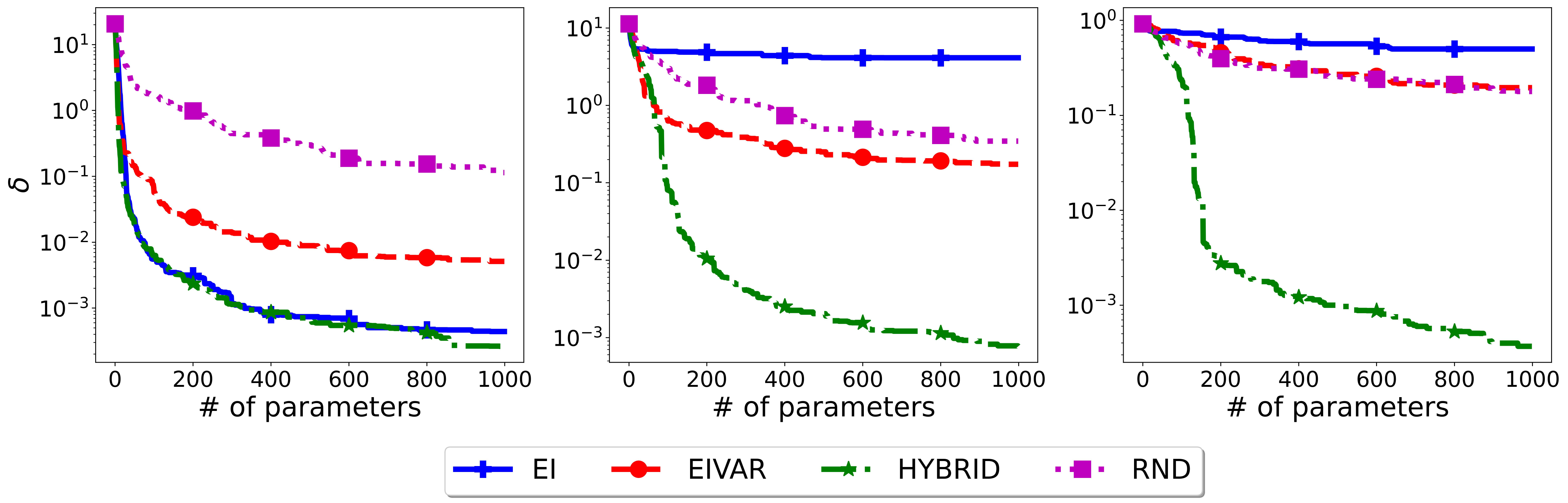}
        \caption{Comparison of different acquisition functions with $b=1$, $w=1$, and $n_0 = 10$ to acquire 1000 parameters. $y$-axis shows $\delta_t = \min\{\left|\y - \model(\thetav_j)\right|, j = 1, \ldots, n_t\}$ after acquiring each set of parameters with EI, EIVAR, HYBRID, and RND for Himmelblau (left), H\"older (middle), and Easom (right) functions.}
        \label{fig:Figure4}
    \end{figure}

Figure~\ref{fig:Figure4} illustrates the absolute difference between the simulation output of the best set of parameters identified up to stage $t$ and the observed data (i.e., $\delta_t = \min\{\left|\y - \model(\thetav_j)\right|, j = 1, \ldots, n_t\}$) for the Himmelblau, H\"older, and Easom functions. In other words, the calibration error is evaluated at each instance when a simulation output is obtained from the set of acquired parameters. Across all three cases, the HYBRID approach consistently identifies parameter sets that most closely match the observed data, outperforming EI, EIVAR, and RND, with the performance gap being particularly notable in the last two examples. 
In the H\"older function example, EI often becomes trapped in one of the high posterior regions and fails to explore the remaining three high posterior regions. For the Easom function, which has a single high posterior region with a relatively small volume compared to the entire parameter space, EI struggles to locate this region and instead remains stuck in local optima. In contrast, the HYBRID approach leverages the exploratory power of EIVAR, which encourages sampling from high posterior regions, and the exploitation focus of EI. As a result, HYBRID effectively targets and exploits high posterior regions, achieving a more balanced and successful exploration-exploitation tradeoff.

    \begin{figure}[ht]
        \includegraphics[width=1\textwidth]{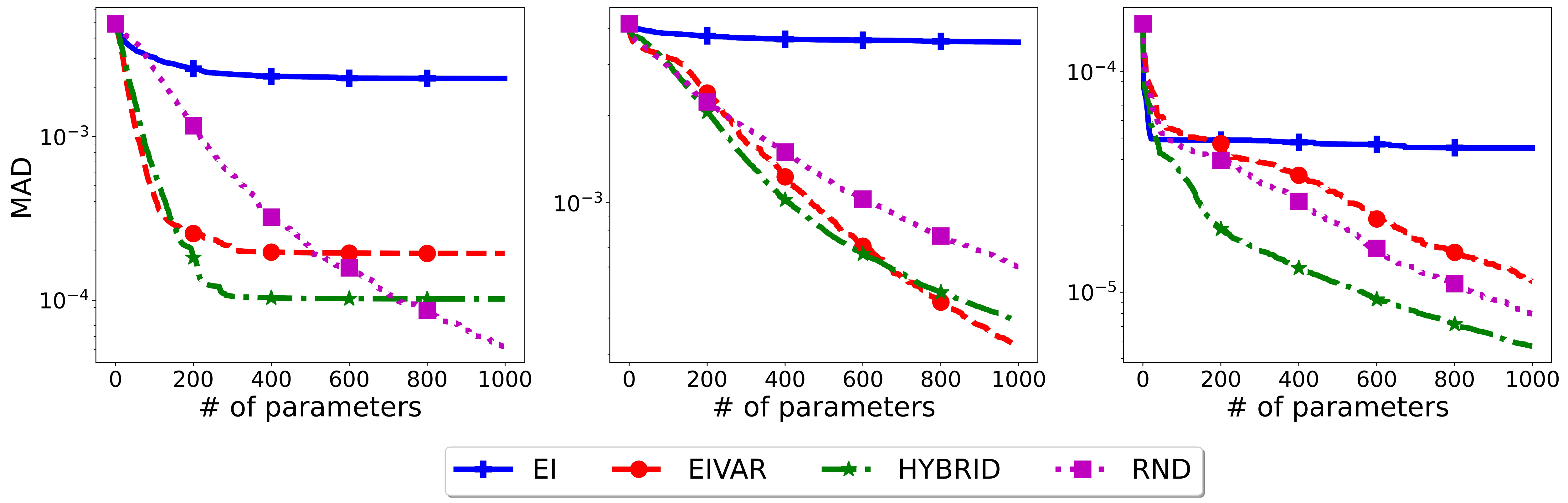}
        \caption{Comparison of different acquisition functions with $b=1$, $w=1$, and $n_0 = 10$ to acquire 1000 parameters. $y$-axis shows ${\rm MAD}_t = \frac{1}{|\Theta_{\rm ref}|} \sum_{\theta \in \Theta_{\rm ref}} |\tilde{p}(\thetav|\y) - \hat{p}_t(\thetav|\y)|$ after acquiring each set of parameters with EI, EIVAR, HYBRID, and RND for Himmelblau (left), H\"older (middle), and Easom (right) functions.}
        \label{fig:Figure5}
    \end{figure}

As discussed in the introduction, calibration can serve multiple purposes. While the primary goal may be to find a simulation output that closely matches the observed data, another objective could be to accurately estimate the posterior density, which in turn leads to more precise parameter inference. An acquisition function that excels in achieving one of these goals may not necessarily be the best for the other, depending on the specific problem. To demonstrate this, we also evaluate how well the posterior is estimated using the emulators constructed with the collected data. We report the mean absolute difference (MAD) between the estimated posterior and the true posterior, as shown in Figure~\ref{fig:Figure5}.
In the experiments, we generate a set of reference parameters $\Theta_{\rm ref}$ and compute the unnormalized posterior density for each $\thetav \in \Theta_{\rm ref}$. The MAD at each stage $t$ is calculated as ${\rm MAD}_t = \frac{1}{|\Theta_{\rm ref}|} \sum_{\theta \in \Theta_{\rm ref}} |\tilde{p}(\thetav|\y) - \hat{p}_t(\thetav|\y)|$, where $\tilde{p}(\thetav|\y)$ is the unnormalized posterior density and $\hat{p}_t(\thetav|\y)$ is the estimated unnormalized posterior at stage $t$, given by $\hat{p}_t(\thetav|\y) = f_\mathcal{N}\left(\y; \, m_t(\thetav), \, \Sigmav + s^2_t(\thetav)\right) p(\thetav)$. 

These results reveal that EI struggles to explore the entire high posterior region, leading to an inaccurate estimation of the overall posterior. The third example is particularly noteworthy, as HYBRID significantly outperforms EIVAR in posterior estimation. In this case, although the posterior has a single peak, the density does not drop to zero across the parameter space. EIVAR tends to select parameters from the entire space, especially at the edges where uncertainty is higher. By alternating between EIVAR and EI through the HYBRID approach, the method better targets the region of interest. Similar results for the Sphere, Matyas, and Ackley functions are provided in Appendix~\ref{app:additionalres}. In all three cases, the posterior density is unimodal, making EI and HYBRID the two most effective acquisition functions for identifying the best parameter set (see Figure~\ref{fig:Figure12}). However, EI falls short in fully exploring the posterior region (see Figure~\ref{fig:Figure13}).

The performance of acquisition functions is influenced by their hyperparameters, which should be carefully selected based on the available computational budget. For instance, with the HYBRID acquisition function, both predictive performance and the time required to acquire a set of parameters are affected by the size of the candidate list $|\mathcal{L}_t|$. To examine the sensitivity of HYBRID to the candidate list size, we vary $|\mathcal{L}_t|$ across three values: 10, 100, and 1000, and then acquire 1000 parameter sets for the Himmelblau function. Using this experimental data, we fit a progress curve for each acquisition function, as shown in Figure~\ref{fig:Figure1}. Here, $\mathcal{A}_1$, $\mathcal{A}_2$, $\mathcal{A}_3$, and $\mathcal{A}_4$ represent HYBRID with candidate sizes of 1000, 100, 10, and RND, respectively. The figure indicates that a larger candidate list leads to longer acquisition times and the accuracy improves as the candidate list size increases. The optimal choice of hyperparameters is the one that balances computational cost with accuracy.

We examine the performance of the HYBRID approach in a parallel setting with varying batch sizes $b \in \{1, 5, 25, 125\}$ using the procedure outlined in Algorithm~\ref{alg:Alg1} and a worker size of $w = 125$. The results for the Sphere, Matyas, and Ackley functions are summarized in Figure~\ref{fig:Figure6}, with additional results for other functions provided in Figure~\ref{fig:Figure14} in Appendix~\ref{app:additionalres}. In this setting, smaller batch sizes allow the GP model to be built from scratch more frequently than the larger batch sizes using the simulation outputs at each stage obtained from acquired parameters. Moreover, with asynchronous approaches a GP model is built with some missing evaluations, whereas all simulation evaluations are complete with $b=125$ before a GP model is built at each stage. On the other hand, with a constant liar strategy, when constructing a batch, the GP model is updated before having to see the simulation outputs. A small batch size strategy can become inefficient as frequent GP model training increases the computational expense. However, the acquisition of multiple sets of parameters at a time with a large batch size leads to the problem of selecting suboptimal sets of parameters since the interaction between the sets of parameters and their outputs within a batch is not successfully taken into account; thus, the quality of the constant liar strategy starts degrading with increasing batch size. As a result, one may need a larger number of runs with larger batch sizes to achieve the same level of accuracy with the smaller batch sizes.
    \begin{figure}[ht]
        \includegraphics[width=1\textwidth]{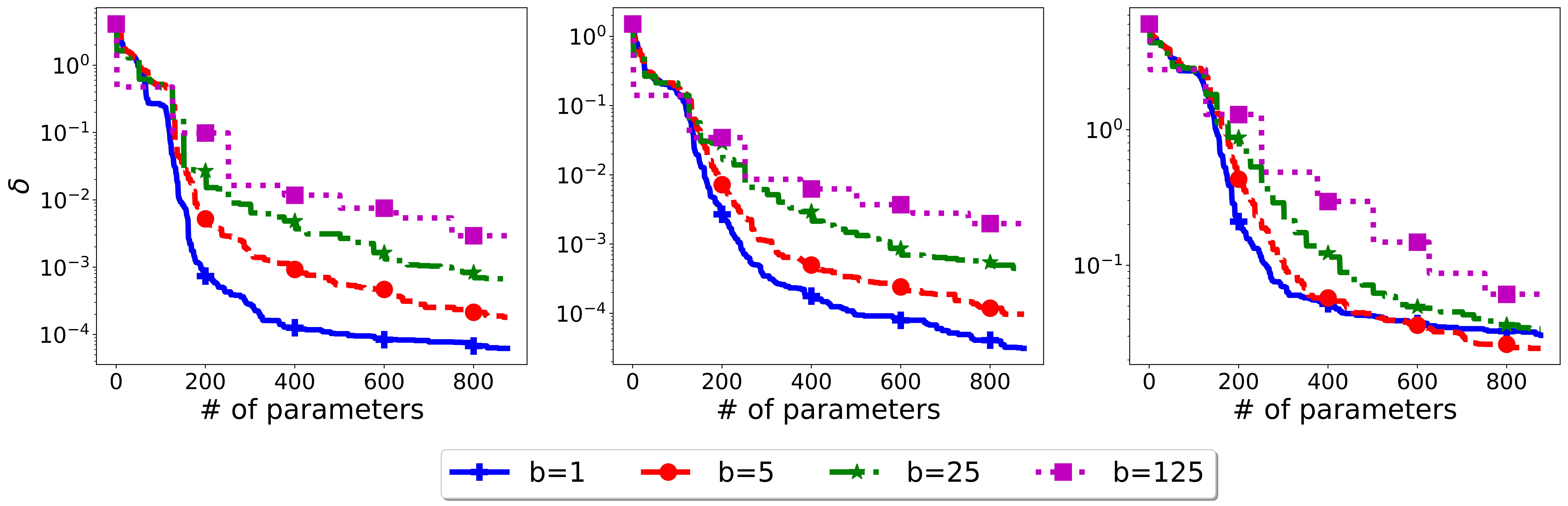}
        \caption{Comparison with $b \in \{1, 5, 25, 125\}$, $w=125$, and $n_0 = 10$ to acquire 1000 parameters via HYBRID for Sphere (left), Matyas (middle), and Ackley (right) functions.}
        \label{fig:Figure6}
    \end{figure}

\subsection{Analysis with Performance Model}
\label{sec:performanceexperiments}

\subsubsection{Code Design.}
 
\texttt{PUQ} includes a Python-based implementation of the proposed acquisition functions, tailored for sequential simulation evaluations. It supports parallel execution on large-scale high-performance computing resources, significantly enhancing computational efficiency. The package includes the \texttt{PUQ.performance} module, which implements the Monte Carlo performance model approach detailed in Algorithm~\ref{alg:Alg2}. The \texttt{performance} module allows users to test with various configurations of batch size $b$ and worker size $w$ in conjunction with different acquisition functions. To describe specific configurations, let $\breve{a}$, $\breve{b}$, and $\breve{c}$ represent input parameters for \texttt{performance} module, as explained further in this section. We use superscripts $n$, $A$, and $S$ to denote parameters related to $n_k(b, \alpha)$, acquisition times, and simulation times, respectively.

\begin{figure}[ht]
\centering
    \begin{subfigure}{1\textwidth}
        \includegraphics[width=1\textwidth]{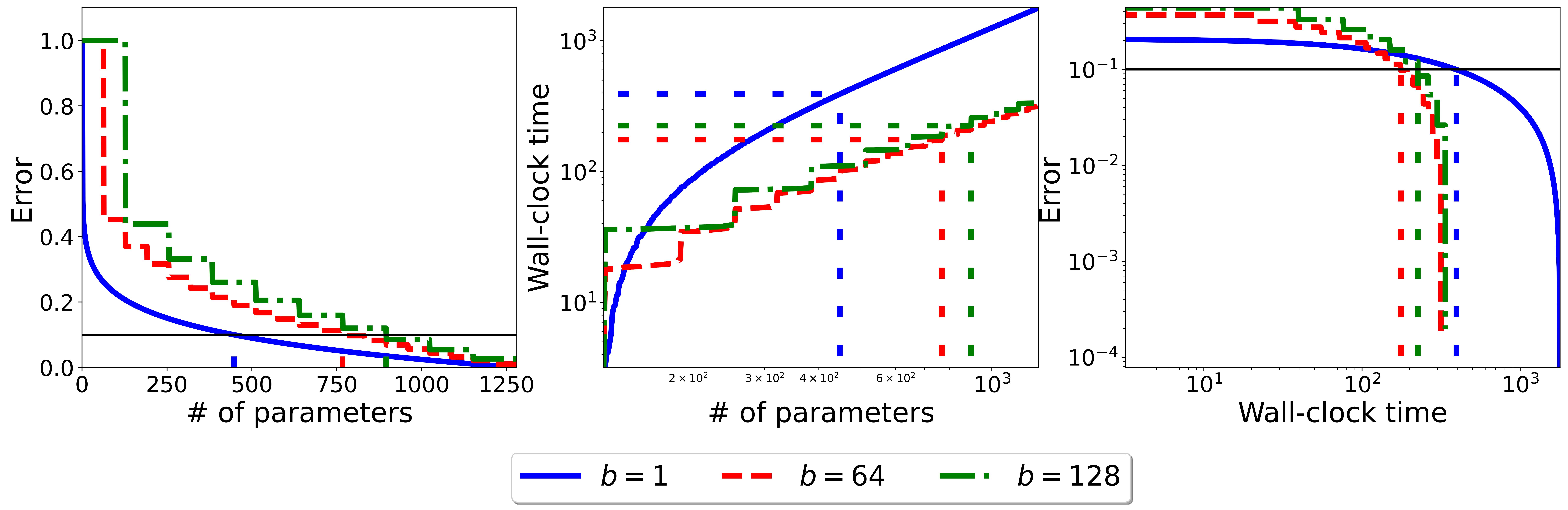}
    \end{subfigure}
    \caption{Comparison of accuracy with different batch sizes $b \in \{1, 64, 128\}$ and worker size $w = 128$. Progress curves are obtained with $\breve{a}^n \in \{0.10, 0.20, 0.25\}$ for $b \in \{1, 64, 128\}$ and $n = 1280$. For each job $j \in \{1 + b(t-1): t = 1, \ldots, \frac{n}{b}\}$, the linear acquisition time with $\breve{a}^A=1$ and $\breve{b}^A=1$ is obtained. For the remaining jobs, the constant acquisition time with $\breve{a}^A = 0.25$ is used. Truncated normal distribution with mean $\breve{a}^S = 1$, standard deviation $\breve{b}^S = 1$, and threshold $\breve{c}^S = 0.1$ is used for simulation run time.}
    \label{fig:Figure7}
\end{figure}

The progress curve is crucial for determining the total number of simulation evaluations, $n_k(b, \alpha)$, needed to achieve a specified accuracy level $\alpha$ with a batch size $b$. Several test cases discussed in Section~\ref{sec:syntheticresults} are available in \texttt{PUQ}, allowing users to apply their chosen acquisition functions and gather data on the number of evaluations and the corresponding error rates.
If prior experimental data is available, users can fit a progress curve based on these observations, as demonstrated in Figure~\ref{fig:Figure1}. This fitted curve can then be used to explore different benchmarking perspectives. In the absence of prior data, users can default to the built-in \texttt{exponential} curve in the \texttt{performance} module, which uses the parameter $\breve{a}^n$ to calculate the error rate as $1 - \left(j/n\right)^{\breve{a}^n}$ for each simulation evaluation $j$ out of $n$ total evaluations. For example, in the left panel of Figure~\ref{fig:Figure7}, the blue line represents the curve generated with $\breve{a}^n=0.1$ and $n = 1280$ for a batch size of $b=1$.
Different parallel settings, including various batch sizes and batching strategies like a constant liar and kriging believer, yield distinct progress curves. In the sensitivity analysis discussed later, for batch sizes greater than one, a progress curve is estimated using a piecewise approximation of the $b=1$ curve, with the number of pieces corresponding to the number of stages for a given $b > 1$. For instance, in Figure~\ref{fig:Figure7} (left panel), 10 stages are required to collect $n = 1280$ simulation evaluations with a batch size of $b=128$. Here, assuming the convergence rate slows with increasing batch size, the progress curve for $b=128$ is estimated using a piecewise approximation of the curve with $\breve{a}^n=0.25$ and $b=1$.

In addition to the convergence rate, the time complexity of each acquisition function varies, allowing for a tradeoff between speed and convergence rate to identify a configuration that balances accuracy and computational efficiency. For benchmarking analysis in \texttt{PUQ}, we provide three options for specifying acquisition time across different parameter sets: \texttt{constant} ($\breve{a}^A$), \texttt{linear} $\left(\breve{a}^A + \breve{b}^A \frac{j}{n}\right)$, and \texttt{quadratic} $\left(\breve{a}^A + \breve{b}^A \frac{j}{n} + \breve{c}^A \left(\frac{j}{n}\right)^2\right)$, where $j$ denotes the parameter index. Algorithm~\ref{alg:Alg2} requires $a_{\omega, k}(b,t)$, the time to acquire $b$ sets of parameters at each stage $t$. This value is the sum of the acquisition times for each of the $b$ sets of parameters generated at that stage. In Algorithm~\ref{alg:Alg1}, an emulator is constructed before acquiring the first set of parameters within a batch, and this same emulator is used throughout the batch's construction. To account for this in our benchmarking analysis, we assign a larger acquisition time for the initial set of parameters generated at each stage (i.e., for each set $j \in \left\{1 + b(t-1): t = 1, \ldots, \frac{n}{b}\right\}$), and a smaller, constant acquisition time for the remaining sets of parameters. 

The average simulation run time and its variability also influence computational demands. The run time for each simulation evaluation can be estimated using built-in \texttt{constant} or truncated \texttt{normal} options. With the \texttt{constant} option, the run time $s_{\omega, j}$ for simulation evaluation $j$ is equal to $\breve{a}^S$, so that $s_{\omega, j} = \breve{a}^S$ for all $j, \omega$. The truncated normal distribution option reflects variation in run times, with $\breve{a}^S$ and $\breve{b}^S$ representing the mean and standard deviation, respectively, and $\breve{c}^S$ serving as the lower bound—if a randomly generated run time falls below $\breve{c}^S$, it is automatically truncated to $\breve{c}^S$.

\subsubsection{Sensitivity Analysis.}
\label{sec:sensitivity}

Figure~\ref{fig:Figure7} illustrates different performance metrics for three batch sizes. As shown in the left panel of Figure~\ref{fig:Figure7}, inspired by the results in Figure~\ref{fig:Figure6}, the acquisition function progresses more slowly with an increasing batch size in terms of the number of simulation evaluations. For example, the number of evaluations required to achieve an error level of $\alpha = 0.1$ is $n_k(1, 0.1) = 447$, $n_k(4, 0.1) = 767$, and $n_k(128, 0.1) = 895$ for $b=1$, $b=4$, and $b=128$, respectively. To provide a more practical measure of efficiency, we consider the total elapsed wall-clock time accounting for acquisitions and $n_k(b, \alpha)$ simulation evaluations. Wall clock time is a widely regarded metric for assessing the performance of parallel code \citep{Barr1993}. In Figure~\ref{fig:Figure7}, the middle and right panels depict the total elapsed wall-clock time as a function of the number of evaluations and the error level, respectively. Although the rate of progress is faster with $b=1$, the sequential approach finishes earlier with $b=64$ due to the smaller acquisition time associated with larger batch sizes.

\begin{figure}[ht]
    \centering
    \includegraphics[width=1\textwidth]{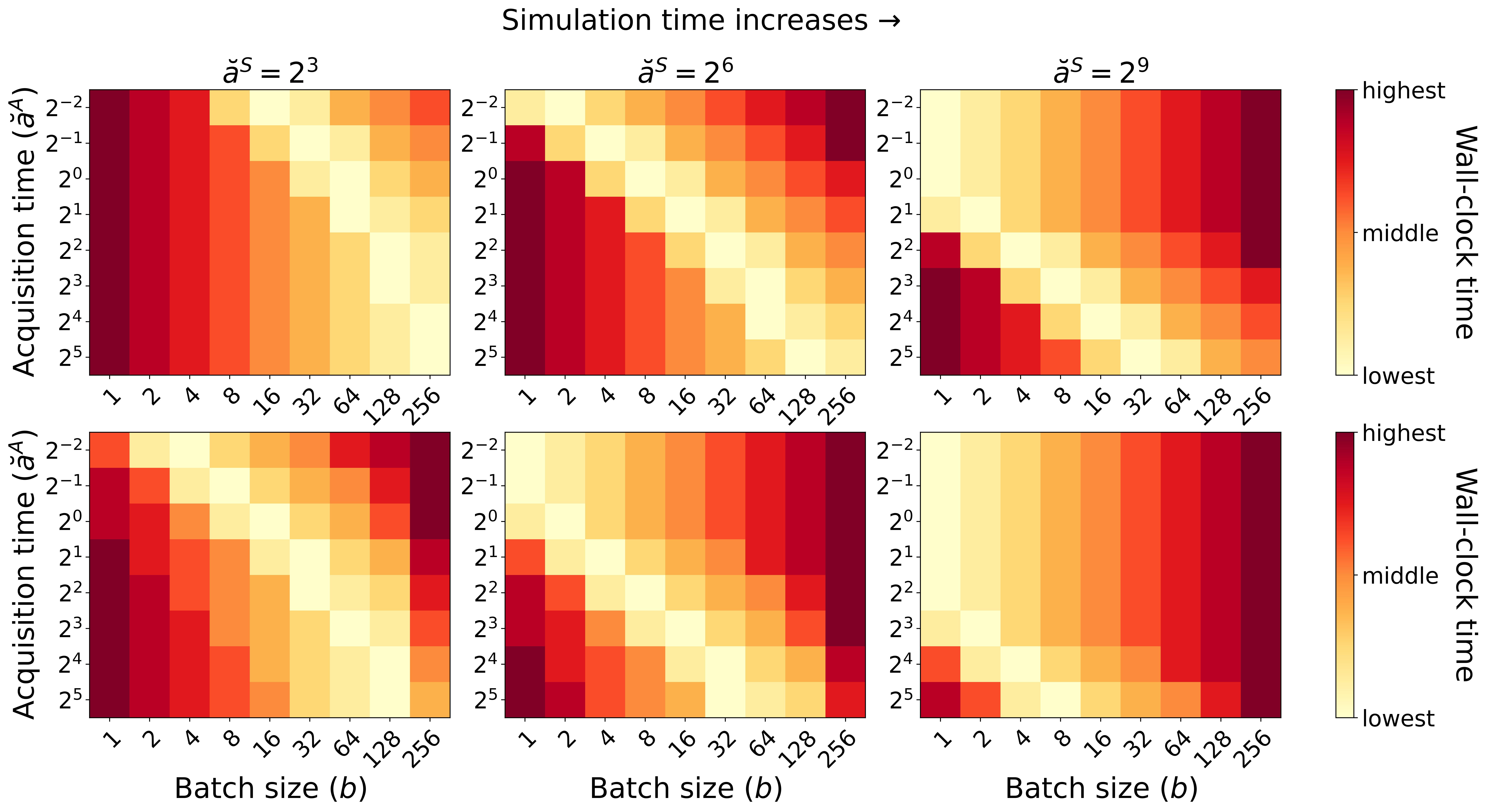}
    \caption{Total elapsed wall-clock time for different batch sizes $b \in \{1, 2, \ldots, 128, 256\}$ to reach an error level of $\alpha = 0.2$ with $w = 256$. Progress curves are obtained with $\breve{a}^n \in \{0.20, 0.22, \ldots, 0.34, 0.36\}$ for $b \in \{1, 2, \ldots, 128, 256\}$ and $n = 2560$. The mean simulation time ($\breve{a}^S$) increases from left panel to right with the simulation time standard deviation set to $\breve{b}^S = 0.1 \times \breve{a}^S$ (first row) and $\breve{b}^S = 10 \times \breve{a}^S$ (second row) and $\breve{c}^S = 0.01$ (obtained with truncated normal distribution). The linear acquisition time with $\breve{a}^A = \breve{b}^A$ is obtained for each job $j \in \{1 + b(t-1): t = 1, \ldots, \frac{n}{b}\}$. For the remaining jobs, the constant acquisition time with $\breve{a}^A = 0.001$ is used.}
    \label{fig:Figure8}
\end{figure}
Figure~\ref{fig:Figure8} presents an extensive study examining the total elapsed wall-clock time across various batch sizes, taking into account both the means and variances of simulation run times and acquisition times. When acquisition time is relatively significantly longer than simulation time (see the left panel of the first row), the synchronous case ($b=256$) outperforms all asynchronous cases. Although asynchronous updates require fewer simulation evaluations to achieve $\alpha$, the synchronous case benefits from fewer stages overall, leading to a shorter total acquisition time. As acquisition time shortens, smaller batch sizes become preferable to achieve a balance between minimizing simulation evaluations and
limiting overall acquisition time. The best batch size changes as simulation times increase (moving from the left panel to the right). In cases where simulation time is relatively significantly longer (as shown in the right panel), a batch size $b=1$ becomes most effective, as shorter acquisition time with more frequent updates outweighs the wait for ongoing jobs to finish when $b > 1$. Increased variability in run times (comparing first-row results with second-row results) further favors smaller batch sizes to minimize the wait time for completed jobs.

\begin{figure}[ht]
\centering
    \begin{subfigure}{1\textwidth}
        \includegraphics[width=1\textwidth]{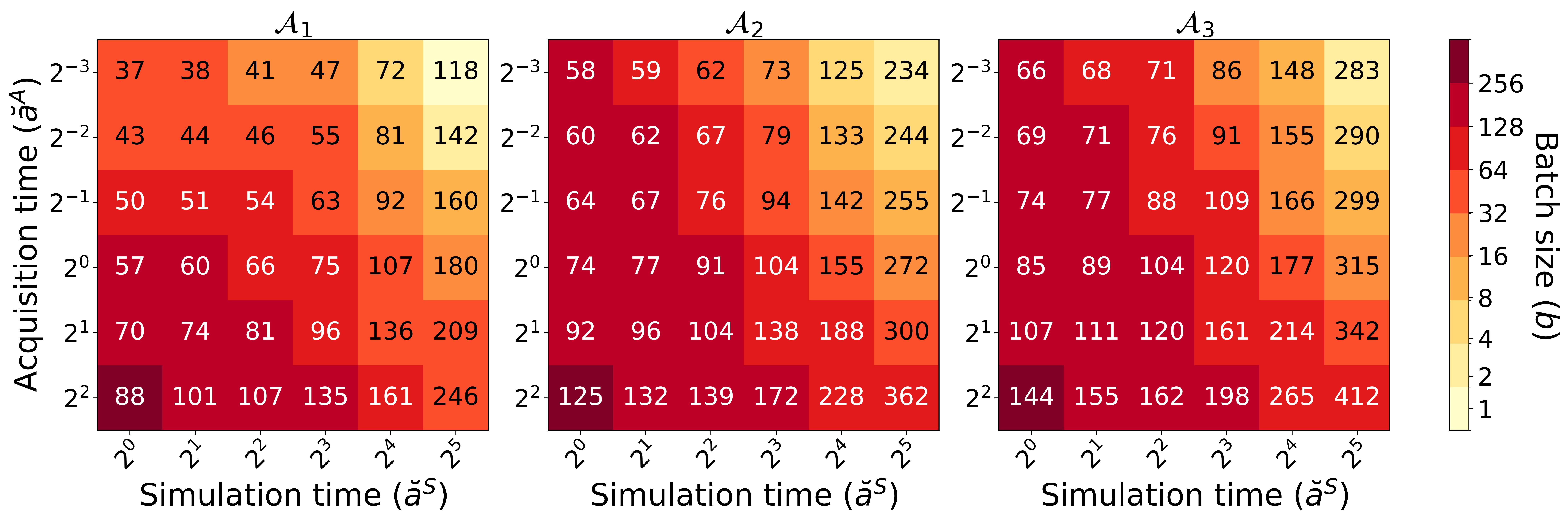}
    \end{subfigure}
    \caption{Batch size with various simulation and acquisition times for acquisition functions $\mathcal{A}_1$, $\mathcal{A}_2$, and $\mathcal{A}_3$ with $w = 256$. Colors indicate the batch size that minimizes the total elapsed wall-clock time to achieve an $\alpha$-level solution and numbers in each cell represent the associated wall-clock time. Progress curves of $\mathcal{A}_1$, $\mathcal{A}_2$, and $\mathcal{A}_3$ are obtained with $\breve{a}^n \in \{0.20, 0.30, 0.40\}$ for $b=1$ and $n=2048$. For $b > 1$, the associated $\breve{a}^n$ value is increased by 0.01 sequentially in the ascending order of $b$. The simulation time is obtained with truncated normal distribution with standard deviation set to $\breve{b}^S = 0.1 \times \breve{a}^S$ and $\breve{c}^S = 0.01$. The linear acquisition time with $\breve{a}^A = \breve{b}^A$ is obtained for each job $j \in \{1 + b(t-1): t = 1, \ldots, \frac{n}{b}\}$. For the remaining jobs, the constant acquisition time with $\breve{a}^A = 0.001$ is used.}
    \label{fig:Figure9}
\end{figure}
As previously discussed, the literature offers several competing acquisition functions, and practitioners must select one for experimentation. In a parallel computing environment, the best batch size for one acquisition function may not be suitable for others, making it crucial to choose the best parallel configuration alongside the acquisition function. Figure~\ref{fig:Figure9} shows the best batch sizes and associated total elapsed wall-clock times required to achieve an error level of $\alpha = 0.2$ for different acquisition functions $\mathcal{A}_1$, $\mathcal{A}_2$, and $\mathcal{A}_3$ across varying simulation and acquisition times. In general, as the ratio of simulation time to acquisition time increases, smaller batch sizes become more favorable for each acquisition function. When acquisition times are equal, the function that requires fewer simulation evaluations (e.g., $\mathcal{A}_1$ in Figure~\ref{fig:Figure9}) outperforms others. However, the time complexity of the acquisition functions often increases with faster error decay rates (e.g., EI vs.\ HYBRID approaches in our experiments). In such cases, one needs to consider the tradeoff between the number of simulation evaluations and the acquisition time. For instance, with simulation time $\breve{a}^S = 2^3$ in Figure~\ref{fig:Figure9}, while $\mathcal{A}_1$ takes 135-time units to reach $\alpha$ with $\breve{a}^A = 2^2$, $\mathcal{A}_2$ becomes preferable if its acquisition time is equal to or less than $\breve{a}^A = 2^0$.

\begin{figure}[ht]
\centering
    \begin{subfigure}{1\textwidth} 
        \includegraphics[width=1\textwidth]{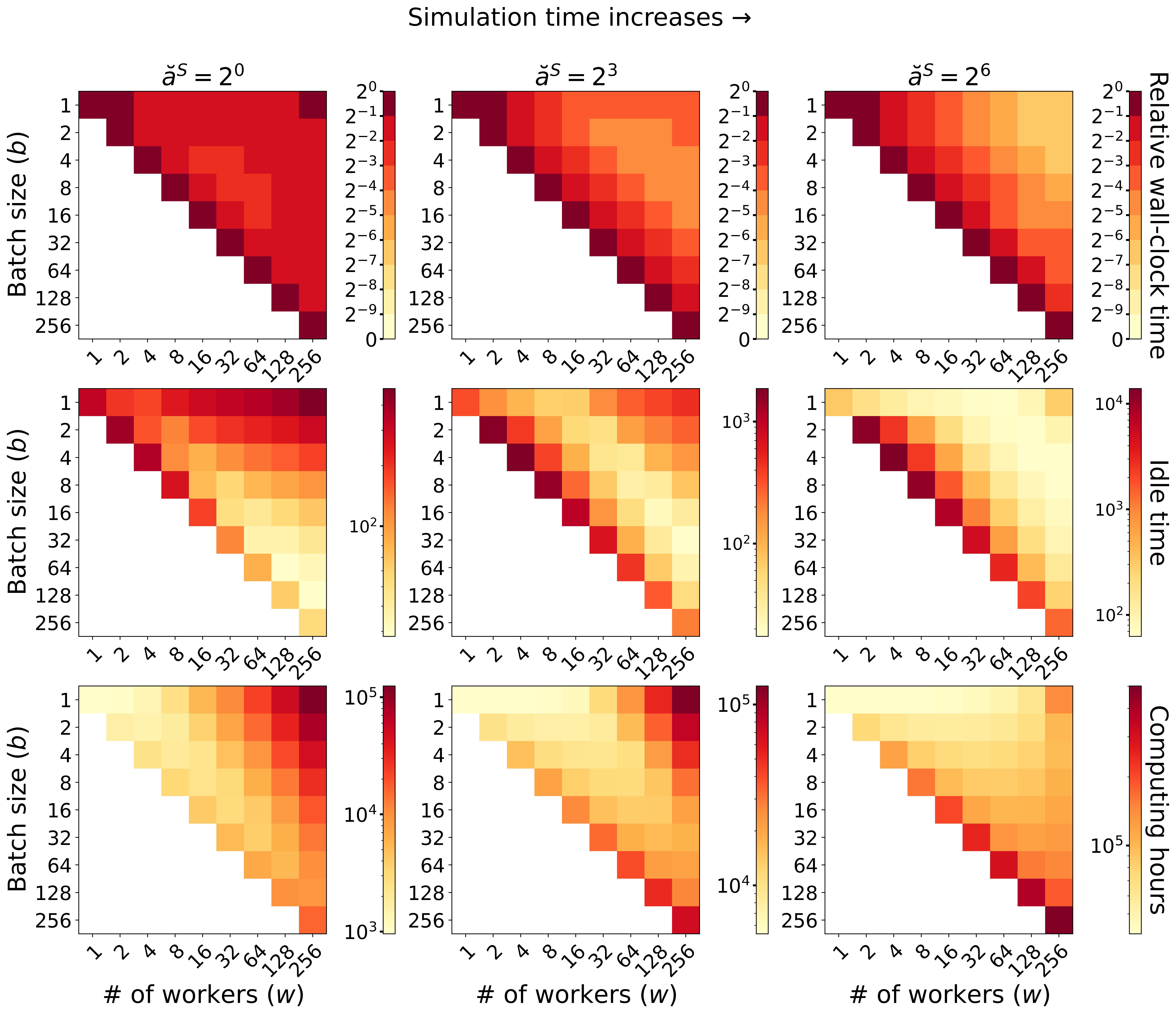}
    \end{subfigure}
    \caption{Analysis with various number of workers $w$ and batch sizes $b$.  The simulation time is obtained with truncated normal distribution with standard deviation set to $\breve{b}^S = \breve{a}^S$ and $\breve{c}^S = 0.01$. The linear acquisition time with $\breve{a}^A = \breve{b}^A$ is obtained for each job $j \in \{1 + b(t-1): t = 1, \ldots, \frac{n}{b}\}$. For the remaining jobs, the constant acquisition time with $\breve{a}^A = 0.001$ is used.}
    \label{fig:Figure10}
\end{figure}
We demonstrate the system's scalability in terms of the number of workers $w$ and the batch size $b$. Figure~\ref{fig:Figure10} highlights three key performance metrics: speedup (first row), average idle time of $w$ workers (second row), and total computing hours (third row). Speedup is a widely used metric in parallel computing environments, with several definitions available \citep{Barr1993}. Typically, it is calculated as the ratio of the time required to solve a problem using one worker to the time required using $w$ workers. However, when the time for $w=1$ is unavailable or the simulation is too expensive to run on a single worker, we compute speedup as the comparison between the execution time for $w$ workers and batch size $b$ and the execution time using the smallest available number of workers for the same batch size.
In the first row of Figure~\ref{fig:Figure10}, the total elapsed wall-clock time for each worker and batch size pair is scaled relative to the total elapsed wall-clock time obtained with the smallest worker size $w=1$ and its corresponding batch size (i.e., the scaled time for worker size 1 equals 1 for each batch size). The second row illustrates the average idle time of $w$ workers, while the third row focuses on energy consumption, calculated as the product of worker size and wall-clock time, since all workers remain occupied until the desired level of accuracy is achieved.

Ideal scaling refers to a parallel configuration that achieves linear scaling, where the time to achieve a solution of a particular quality decreases proportionally to the number of workers used. In other words, doubling the number of parallel workers would halve the wall-clock time. However, in the left panel, where the simulation time is relatively short, the wall-clock time demonstrates poor scaling behavior. For $b=1$, the worst performance occurs with $w=256$. As shown by the idle times in the second row, while one worker performs the acquisitions, the remaining workers are idle, waiting for the acquisitions to complete. This results in excessive computing hour consumption. As simulation time increases (from left to right), smaller batch sizes tend to perform better for a given worker size $w$ in terms of all three metrics. The relative wall-clock time approaches the ideal scaling for the longest simulation time. In these scenarios, where simulation times dominate, acquisitions occur concurrently with parameter evaluations in the simulation model.
Users often have a limited budget for computing hours on large high-performance computing resources, so considering energy consumption when configuring the experimental design is essential. Understanding the performance characteristics of the sequential procedure and their impact on energy usage is crucial for effective computational resource planning.

%acquisition time becomes a significant portion of the wall-clock time needed to reach the desired error level in these cases.

\section{Conclusion}

We propose a new performance model to assess numerous performance characteristics of the sequential design in a parallel environment for calibration and provide an associated toolset implemented in the PUQ software package. The benefits of our toolset include a cheaper identification of effective design in a parallel setting as well as an improved understanding of how performance varies with the characteristics of a simulation model and various acquisition functions. The proposed performance model can be used to test and benchmark alternative regimes. For example, one alternative is to dynamically decide the batch size at each stage similar to that of \cite{Azimi2010} rather than using a fixed batch size until the procedure is terminated as illustrated in this paper. Throughout the paper, we use non-preemptive allocation of jobs where each worker is occupied until a simulation evaluation is completed. The performance model can be further extended for the preemptive allocation of jobs to utilize the computational resources better. We demonstrate our performance results with a novel acquisition function that switches between EI and EIVAR criteria to trade off exploitation and exploration for calibration. The proposed acquisition strategy is able to identify the best set of parameters that aligns simulation output with the observed data by reducing the total uncertainty on the high posterior regions. Further enhancement on the hybrid acquisition function is to automatically tradeoff between exploration and exploitation such as the one proposed by \cite{Andrad2009} rather than using EI and EIVAR consecutively. Our toolset is modular so that built-in acquisition functions, the sequential framework, and simulation models can be replaced with user-specific functions. That way, users can gather data to gain insight into the performance characteristics of their sequential procedure as well, and comparisons between alternative procedures promote innovation in research. In this work, we assume uniform variability in simulation time across the parameter space. However, simulation time can depend on the parameter space, e.g., being shorter and less variable in high posterior regions compared to other regions. Analyzing performance under this setting would be an interesting direction for further study.

% Acknowledgments here
\ACKNOWLEDGMENT{%
    This work was supported by the NSF CSSI program under grant \rm{OAC-2004601} (BAND Collaboration) and by the U.S.~Department of Energy,
Office of Science, Office of Advanced Scientific Computing Research's applied mathematics and SciDAC
programs under Contract No.\ DE-AC02-05CH11231. Generous computing resources provided on Bebop, a high-performance computing cluster operated by the Laboratory Computing Resource Center at Argonne National Laboratory, are gratefully acknowledged.
}% Leave this (end of acknowledgment)

% Appendix here
% Options are (1) APPENDIX (with or without general title) or 
%             (2) APPENDICES (if it has more than one unrelated sections)
% Outcomment the appropriate case if necessary
%
% \begin{APPENDIX}{<Title of the Appendix>}
% \end{APPENDIX}
%
%   or 
%
\newpage
\begin{APPENDICES}
    \section{Proofs}
\label{app:proofs}

\begin{proposition}\label{prop:EIVAR:stochastic}
The expected probability of improvement for $\thetav^*$ is given by
    \begin{align} \label{eq:EIVcriterion}
        \begin{split}
            \Phi\left(\frac{\delta_t - \left(y - m_t(\thetav^*)\right)}{\sigma^2 + s_t^2\left(\thetav^*\right)}\right) -
            \Phi\left(\frac{- \delta_t - (y - m_t(\thetav^*))}{\sigma^2 + s_t^2\left(\thetav^*\right)}\right),
        \end{split}
    \end{align}
where $\delta_t = \min\{|\y - \model(\thetav_j)|, j = 1, \ldots, n_t\}$ and $\Phi$ represents the CDF of a standard normal random variable.
\end{proposition}

\proof{}
Due to the error model in \eqref{eq:statmodel}, we can write $\epsilon^* \sim \mathcal{N}\left(y - \model(\thetav^*), \sigma^2\right)$ for any fixed $\model(\thetav^*)$. GP model introduced in Section~\ref{sec:review} assumes $\model(\thetav^*) \sim \mathcal{N}\left(m_t\left(\thetav^*\right), s_t^2\left(\thetav^*\right)\right).$ The expected probability of improvement is then defined as $\mathbb{E}_{\model(\thetav^*)|\mathcal{D}_{t}} \left[\mathbb{P}\left(|\epsilon^*| \leq \delta_t\right)\right]$, which can be equivalently written as
\begin{align} \label{eq:proimpr1}
    \begin{split}
        &=\mathbb{E}_{\model(\thetav^*)|\mathcal{D}_{t}} \left[\mathbb{P}\left(\epsilon^* \leq \delta_t\right) - \mathbb{P}\left(\epsilon^* \leq -\delta_t\right)\right] \\
        &= \mathbb{E}_{\model(\thetav^*)|\mathcal{D}_{t}} \left[\mathbb{P}\left(\frac{\epsilon^*-\left(\y - \model(\thetav^*)\right)}{\sigma} \leq \frac{\delta_t-\left(\y - \model(\thetav^*)\right)}{\sigma}\right) - \mathbb{P}\left(\frac{\epsilon^*-\left(\y - \model(\thetav^*)\right)}{\sigma} \leq \frac{-\delta_t-\left(\y - \model(\thetav^*)\right)}{\sigma}\right)\right] \\
        &= \int \Phi\left(\frac{\delta_t-\left(\y - \model(\thetav^*)\right)}{\sigma}\right) \pdfN{\model(\thetav^*)}{m_t(\thetav^*)}{s^2_t(\thetav^*)} d\model(\thetav^*) \\ &-
        \int \Phi\left(\frac{-\delta_t-\left(\y - \model(\thetav^*)\right)}{\sigma}\right) \pdfN{\model(\thetav^*)}{m_t(\thetav^*)}{s^2_t(\thetav^*)} d\model(\thetav^*).
    \end{split}
\end{align}
By changing the variables, let $\model(\thetav^*) = m_t(\thetav^*) + s_t(\thetav^*) z$ where $z \sim \mathcal{N}(0, 1)$. Then, \eqref{eq:proimpr1} is equivalently written
\begin{align} \label{eq:proimpr2}
    \begin{split}
        &\int \Phi\left(\frac{\delta_t-\left(\y - m_t(\thetav^*) - s_t(\thetav^*) z\right)}{\sigma}\right) \phi(z) dz -
        \int \Phi\left(\frac{-\delta_t-\left(\y - m_t(\thetav^*) - s_t(\thetav^*) z\right)}{\sigma}\right) \phi(z) dz.
    \end{split}
\end{align}
Using $\int_{-\infty}^{\infty} \Phi(a + bx) \phi(x) dx = \Phi\left(\frac{a}{\sqrt{1 + b^2}}\right)$, \eqref{eq:proimpr2} is equivalently written as \eqref{eq:EIVcriterion}.

\begin{proposition}
The expected unimprovement for $\thetav^*$ is given by
    \begin{align} \label{eq:expectedimprovement_app}
        \begin{split}
            & \left(\y - m_t(\thetav^*) - \delta_t\right)  \left(1 - \Phi\left(\frac{\delta_t - \left(\y - m_t(\thetav^*)\right)}{\sqrt{\sigma^2 + s_t^2\left(\thetav^*\right)}}\right)\right) + \sqrt{\sigma^2 + s_t^2\left(\thetav^*\right)} \phi\left(\frac{\delta_t -\left( \y - m_t(\thetav^*)\right)}{\sqrt{\sigma^2 + s_t^2\left(\thetav^*\right)}}\right)  \\
            & + \left(-\left(\y - m_t(\thetav^*)\right) - \delta_t\right) \Phi\left(\frac{-\delta_t - \left(\y - m_t(\thetav^*)\right)}{\sqrt{\sigma^2 + s_t^2\left(\thetav^*\right)}}\right) + \sqrt{\sigma^2 + s_t^2\left(\thetav^*\right)}\phi\left(\frac{-\delta_t -\left( \y - m_t(\thetav^*)\right)}{\sqrt{\sigma^2 + s_t^2\left(\thetav^*\right)}}\right) 
        \end{split}
    \end{align}
where $\delta_t = \min\{\left|\y - \model(\thetav_j)\right|, j = 1, \ldots, n_t\}$ and $\Phi$ represents the CDF of a standard normal random variable.
\end{proposition}
\proof{}
Due to the error model in \eqref{eq:statmodel}, we can write $\epsilon^* \sim \mathcal{N}\left(y - \model(\thetav^*), \sigma^2\right)$ for any fixed $\model(\thetav^*)$. GP model introduced in Section~\ref{sec:review} assumes $\model(\thetav^*) \sim \mathcal{N}\left(m_t\left(\thetav^*\right), s_t^2\left(\thetav^*\right)\right).$ The expected unimprovement is then defined as $\mathbb{E}_{\model(\thetav^*)|\mathcal{D}_{t}}\left[\max\left(\epsilon^* - \delta_t, 0\right)\right] + \mathbb{E}_{\model(\thetav^*)|\mathcal{D}_{t}}\left[\max\left(-\epsilon^* - \delta_t, 0\right)\right]$, which can be equivalently written as
\begin{align} \label{eq:expectedimprovement1}
    \begin{split}
        &= \int_{-\infty}^{\infty} \left(\int_{\delta_t}^{\infty} (\epsilon^* - \delta_t)  \pdfN{\epsilon^*}{y - \model(\thetav^*)}{\sigma^2} d\epsilon^* \right)\pdfN{\model(\thetav^*)}{m_t(\thetav^*)}{s_t^2\left(\thetav^*\right)} d\model(\thetav^*) \\
        &+ \int_{-\infty}^{\infty} \left(\int_{\infty}^{-\delta_t} (-\epsilon^* - \delta_t)  \pdfN{\epsilon^*}{y - \model(\thetav^*)}{\sigma^2} d\epsilon^* \right) \pdfN{\model(\thetav^*)}{m_t(\thetav^*)}{s_t^2\left(\thetav^*\right)} d\model(\thetav^*). \\
    \end{split}
\end{align}
By changing the variables, let $\epsilon^* = \left(\y - \model(\thetav^*)\right) + \sigma z$ where $z \sim \mathcal{N}(0, 1)$. For the sake of brevity, we drop $\model(\thetav^*)$, $m_t(\thetav^*)$, and $s_t^2\left(\thetav^*\right)$ from $\pdfN{\model(\thetav^*)}{m_t(\thetav^*)}{s_t^2\left(\thetav^*\right)}$. Then, \eqref{eq:expectedimprovement1} is equivalently written
\begin{equation} \label{eq:expectedimprovement2}
    \begin{gathered}
         = \int_{-\infty}^{\infty} \left(\int_{\frac{\delta_t - (\y - \model(\thetav^*))}{\sigma}}^{\infty} ((\y - \model(\thetav^*)) + \sigma z - \delta_t) \phi(z) dz \right) f_{\mathcal{N}} d\model(\thetav^*)  \\
        + \int_{-\infty}^{\infty} \left(\int_{\infty}^{\frac{-\delta_t - (\y - \model(\thetav^*))}{\sigma}} (-((\y - \model(\thetav^*)) + \sigma z) - \delta_t) \phi(z) dz \right) f_{\mathcal{N}} d\model(\thetav^*) \\
         = \int_{-\infty}^{\infty} \left((\y - \model(\thetav^*) - \delta_t)  \left(1 - \Phi\left(\frac{\delta_t - (\y - \model(\thetav^*))}{\sigma}\right)\right) + \sigma \int_{\frac{\delta_t-(\y - \model(\thetav^*))}{\sigma}}^{\infty} z \phi(z) dz\right) f_{\mathcal{N}} d\model(\thetav^*)\\
         + \int_{-\infty}^{\infty} \left((-(\y - \model(\thetav^*)) -\delta_t) \Phi\left(\frac{-\delta_t - (\y - \model(\thetav^*))}{\sigma^2}\right) - \sigma \int_{-\infty}^{\frac{-\delta_t-(\y - \model(\thetav^*))}{\sigma}} z \phi(z) dz\right) f_{\mathcal{N}} d\model(\thetav^*) \\
         = \int_{-\infty}^{\infty} \left((\y - \model(\thetav^*) - \delta_t)  \left(1 - \Phi\left(\frac{\delta_t - (\y - \model(\thetav^*))}{\sigma}\right)\right) + \sigma \phi\left(\frac{\delta_t - (\y - \model(\thetav^*))}{\sigma}\right) \right) f_{\mathcal{N}} d\model(\thetav^*)\\
         + \int_{-\infty}^{\infty} \left((-(\y - \model(\thetav^*)) -\delta_t) \Phi\left(\frac{-\delta_t - (\y - \model(\thetav^*))}{\sigma^2}\right) + \sigma \phi\left(\frac{-\delta_t - (\y - \model(\thetav^*))}{\sigma}\right) \right) f_{\mathcal{N}} d\model(\thetav^*).\\
    \end{gathered}
\end{equation}
By changing the variables, let $\model(\thetav^*) = m_t(\thetav^*) + s_t(\thetav^*) z$ where $z \sim \mathcal{N}(0, 1)$. Then,  
\begin{equation} \label{eq:expectedimprovement3}
    \begin{gathered}
         = \int_{-\infty}^{\infty} \left(\left(\y - m_t(\thetav^*) - \delta_t - s_t(\thetav^*) z\right) \left(1 - \Phi\left(\frac{\delta_t - \left(\y - \left(m_t(\thetav^*) + s_t(\thetav^*) z\right)\right)}{\sigma}\right)\right)\right) \phi(z) dz \\ 
         + \sigma \int_{-\infty}^{\infty} \phi\left(\frac{\delta_t - \left(\y - \left(m_t(\thetav^*) + s_t(\thetav^*) z\right)\right)}{\sigma}\right) \phi(z) dz\\
         + \int_{-\infty}^{\infty} \left(\left(m_t(\thetav^*) - \y - \delta_t + s_t(\thetav^*) z\right)\Phi\left(\frac{-\delta_t - \left(\y - m_t(\thetav^*) - s_t(\thetav^*) z\right)}{\sigma}\right)\right) \phi(z) dz \\ 
         + \sigma \int_{-\infty}^{\infty} \phi\left(\frac{-\delta_t - (\y - (m_t(\thetav^*) + s_t(\thetav^*) z))}{\sigma}\right) \phi(z) dz.
    \end{gathered}
\end{equation}
Using $\int_{-\infty}^{\infty} \Phi(a + bx) \phi(x) dx = \Phi\left(\frac{a}{\sqrt{1 + b^2}}\right)$ and $\int_{-\infty}^{\infty} \phi(a + bx) \phi(x) dx = \frac{1}{\sqrt{1 + b^2}} \phi\left(\frac{a}{\sqrt{1 + b^2}}\right)$, we obtain
\begin{equation} \label{eq:expectedimprovement4}
    \begin{gathered}
         = \left(\left(\y -  m_t(\thetav^*) - \delta_t\right) \left(1 - \Phi\left(\frac{\delta_t - \left(\y - m_t(\thetav^*)\right)}{\sqrt{\sigma^2 + s_t^2\left(\thetav^*\right)}}\right)\right)\right)+ \frac{\sigma^2}{\sqrt{\sigma^2 + s_t^2\left(\thetav^*\right)}} \phi\left(\frac{\delta_t - \left( \y - m_t(\thetav^*)\right)}{\sqrt{\sigma^2 + s_t^2\left(\thetav^*\right)}}\right)\\
         - s_t(\thetav^*)\int_{-\infty}^{\infty} z \left(1 - \Phi\left(\frac{\delta_t - \left(\y - \left(m_t(\thetav^*) + s_t(\thetav^*) z\right)\right)}{\sigma}\right)\right) \phi(z) dz \\ 
         + \left(\left(-(\y - m_t(\thetav^*)) - \delta_t \right) \Phi\left(\frac{-\delta_t - \left(\y - m_t(\thetav^*)\right)}{\sqrt{\sigma^2 + s_t^2\left(\thetav^*\right)}}\right)\right) + \frac{\sigma^2}{\sqrt{\sigma^2 + s_t^2\left(\thetav^*\right)}} \phi\left(\frac{-\delta_t -\left( \y - m_t(\thetav^*)\right)}{\sqrt{\sigma^2 + s_t^2\left(\thetav^*\right)}}\right)\\
         + s_t(\thetav^*)\int_{-\infty}^{\infty} z\Phi\left(\frac{-\delta_t - \left(\y - \left(m_t(\thetav^*) + s_t(\thetav^*) z\right)\right)}{\sigma}\right) \phi(z) dz. \\ 
    \end{gathered}
\end{equation}

Using $\int_{-\infty}^{\infty} x \Phi\left(a + bx\right) \phi(x) dx = \frac{b}{\sqrt{1 + b^2}} \phi\left(\frac{a}{\sqrt{1 + b^2}}\right)$, we finally obtain \eqref{eq:expectedimprovement_app}.

\endproof

\section{Experiment Setting Details}
\label{app:setting}
In this section, we provide the details of the test functions used in our numerical experiments. The test functions are commonly used synthetic functions for optimization (see, e.g., \citep{synthlinks}), and Figure~\ref{synth_figs} illustrates the posterior density for each synthetic function. 
    \begin{figure}[ht]
        \begin{subfigure}{0.33\textwidth}
            \includegraphics[width=1\textwidth]{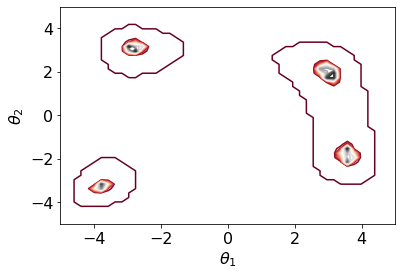}
            \caption{Himmelblau's function}
        \end{subfigure}
        \begin{subfigure}{0.33\textwidth}
            \includegraphics[width=1\textwidth]{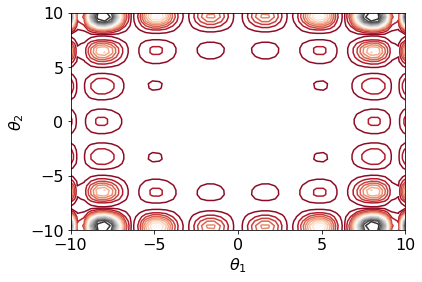}
            \caption{H\"older table function}
        \end{subfigure}
        \begin{subfigure}{0.33\textwidth}
            \includegraphics[width=1\textwidth]{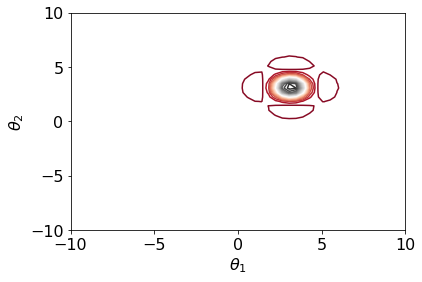}
            \caption{Easom function}
        \end{subfigure}
        \begin{subfigure}{0.33\textwidth}
            \includegraphics[width=1\textwidth]{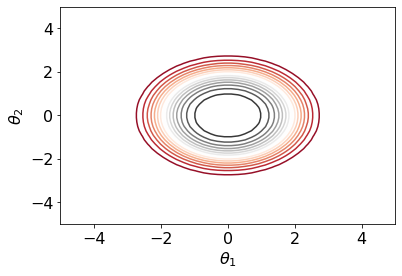}
            \caption{Sphere function}
        \end{subfigure}
        \begin{subfigure}{0.33\textwidth}
            \includegraphics[width=1\textwidth]{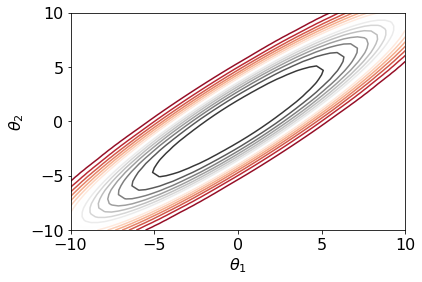}
            \caption{Matyas function}
        \end{subfigure}
        \begin{subfigure}{0.33\textwidth}
            \includegraphics[width=1\textwidth]{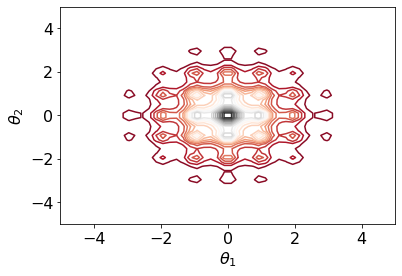}
            \caption{Ackley Function}
        \end{subfigure}        
        \caption{Illustration of six synthetic densities.}
        \label{synth_figs}
    \end{figure}
    
The details of each setting are given as follows in the calibration context:
\begin{itemize}
    \item For Himmelblau's function, we consider $\theta_1 \in [-5, 5]$ and $\theta_2 \in [-5, 5]$. In the calibration context, we take $\thetav = (\theta_1, \theta_2)$ and $\model(\thetav) = (\theta_1^2 + \theta_2 - 11)^2 + (\theta_1 + \theta_2^2 - 7)^2$ with $\y = 1$ and $\Sigmav=1$.
    \item For H\"older table function, we consider $\theta_1 \in [-10, 10]$ and $\theta_2 \in [-10, 10]$. In the calibration context, we take $\thetav = (\theta_1, \theta_2)$ and $\model(\thetav) =  -|\sin(\theta_1) \times \cos(\theta_2) \times \exp(|1 - (\sqrt{\theta_1^2 + \theta_2^2}/\pi)|)|$ with $\y = -19.2085$ and $\Sigmav=50$.
    \item For Easom function, we consider $\theta_1 \in [-10, 10]$ and $\theta_2 \in [-10, 10]$. In the calibration context, we take $\thetav = (\theta_1, \theta_2)$ and $\model\left(\thetav\right) = -\cos\left(\theta_1\right) \cos\left(\theta_2\right) \exp\left(-((\theta_1 - \pi)^2 + (\theta_2 - \pi)^2)\right)$ with $y = -1$ and $\sigma^2 = 10$.
    \item For the sphere function, we consider $\theta_1 \in [-5, 5]$ and $\theta_2 \in [-5, 5]$. In the calibration context, we take $\thetav = (\theta_1, \theta_2)$ and $\model(\thetav) = \theta_1^2 + \theta_2^2$ with $\y = 0$ and $\Sigmav = 10$.
    \item For Matyas function, we consider $\theta_1 \in [-10, 10]$ and $\theta_2 \in [-10, 10]$, $\model(\thetav) = 0.26 (\theta_1^2 + \theta_2^2) - 0.48 \theta_1 \theta_2$ with $\y = 0$ and $\Sigmav = 10$.
    \item For Ackley function, we consider $\theta_1 \in [-5, 5]$ and $\theta_2 \in [-5, 5]$, $\model(\thetav) = -20 \exp\left(-0.2  \sqrt{0.5  (\theta_1^2 + \theta_2^2)}\right) - \exp\left(0.5 \left(\cos\left(2 \pi \theta_1\right) + \cos\left(2 \pi  \theta_2\right)\right)\right) + e + 20
 $ with $\y = 0$ and $\Sigmav = 10$.
\end{itemize}

\section{Additional Experiment Results}
\label{app:additionalres}

Additional results illustrating different acquisition functions are provided in Figures~\ref{fig:Figure12} and \ref{fig:Figure13}. Figure~\ref{fig:Figure14} shows the effects of varying the batch size.

    \begin{figure}[ht]
        \includegraphics[width=1\textwidth]{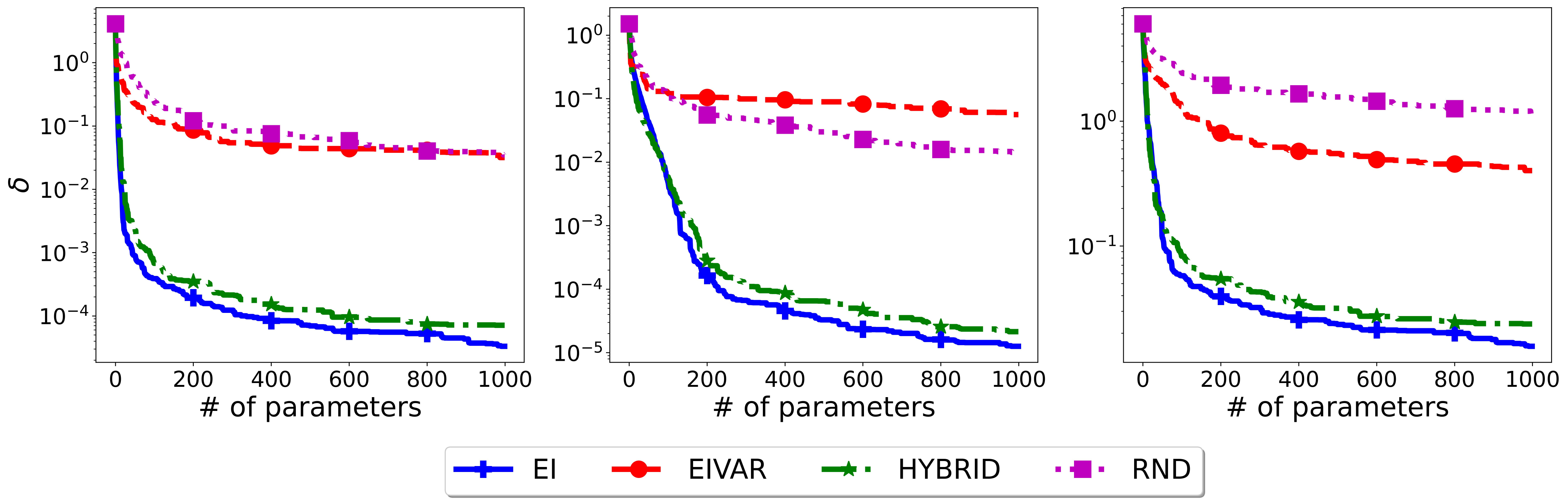}
        \caption{Comparison of different acquisition functions with $b=1$, $w=1$, and $n_0 = 10$ to acquire 1000 parameters. $y$-axis shows $\delta_t = \min\{\left|\y - \model(\thetav_j)\right|, j = 1, \ldots, n_t\}$ after acquiring each set of parameters with EI, EIVAR, HYBRID, and RND for Sphere (left), Matyas (middle), and Ackley (right) functions.}
        \label{fig:Figure12}
    \end{figure}

    \begin{figure}[ht]
        \includegraphics[width=1\textwidth]{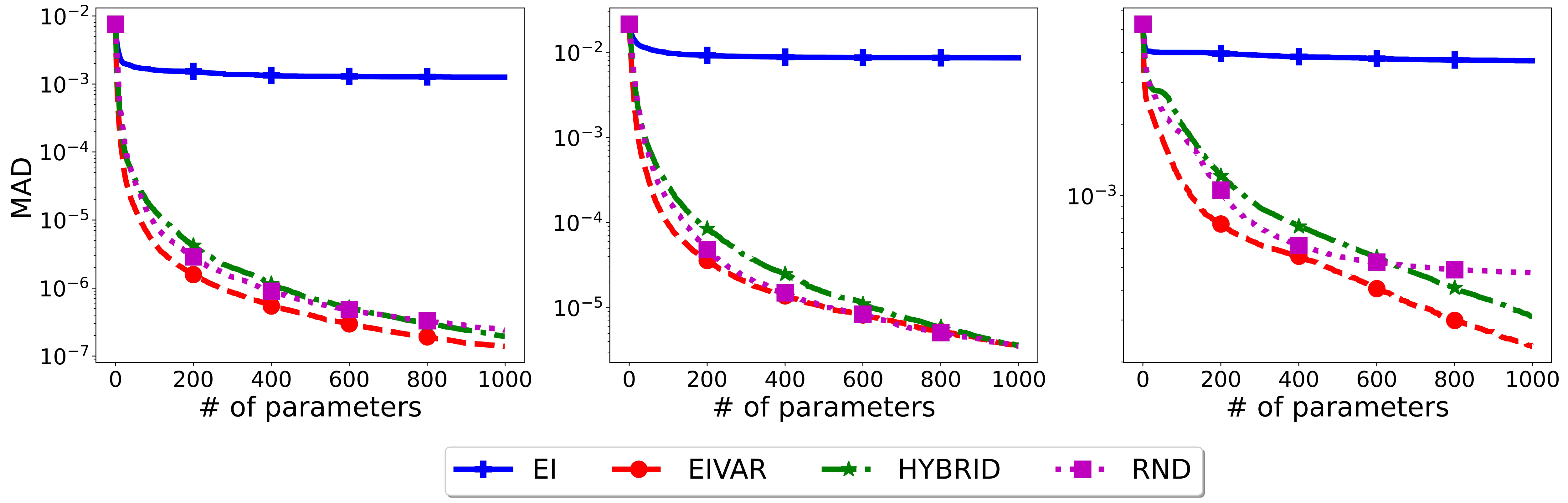}
        \caption{Comparison of different acquisition functions with $b=1$, $w=1$, and $n_0 = 10$ to acquire 1000 parameters. $y$-axis shows ${\rm MAD}_t(\tilde{p}, \hat{p}) = \frac{1}{|\Theta_{\rm ref}|} \sum_{\theta \in \Theta_{\rm ref}} |\tilde{p}(\thetav|\y) - \hat{p}_t(\thetav|\y)|$ after acquiring each set of parameters with EI, EIVAR, HYBRID, and RND for Sphere (left), Matyas (middle), and Ackley (right) functions.}
        \label{fig:Figure13}
    \end{figure}
    
    \begin{figure}[ht]
        \includegraphics[width=1\textwidth]{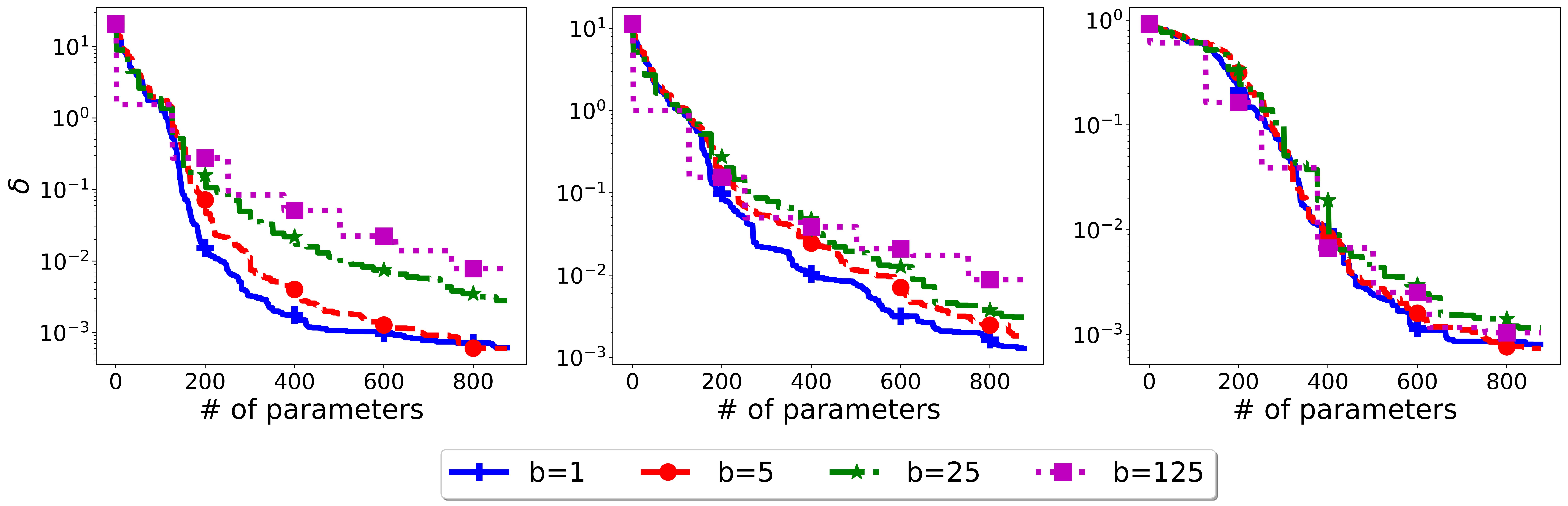}
        \caption{Comparison with $b \in \{1, 5, 25, 125\}$, $w=125$, and $n_0 = 10$ to acquire 1000 parameters via HYBRID for Himmelblau (left), H\"older (middle), and Easom (right) functions.}
        \label{fig:Figure14}
    \end{figure}

\end{APPENDICES}

% References here (outcomment the appropriate case) 

% CASE 1: BiBTeX used to constantly update the references 
%   (while the paper is being written).
\bibliographystyle{informs2014} % outcomment this and next line in Case 1
\bibliography{activelearning} % if more than one, comma separated

% CASE 2: BiBTeX used to generate mypaper.bbl (to be further fine tuned)
%\input{mypaper.bbl} % outcomment this line in Case 2

\end{document}